\documentclass[pra,aps,twocolumn]{revtex4-1}
\usepackage{graphicx}
\usepackage{amsmath}
\usepackage{bm}

\newcommand{\be}{\begin{equation}}
\newcommand{\ee}{\end{equation}}
\newcommand{\bea}{\begin{eqnarray}}
\newcommand{\eea}{\end{eqnarray}}
\newcommand{\bei}{\begin{itemize}}
\newcommand{\eei}{\end{itemize}}

\def\ket#1{\left|{#1}\right>}

\let\oldhat\hat
\renewcommand{\vec}[1]{\mathbf{#1}}
\renewcommand{\hat}[1]{\oldhat{#1}}

\begin{document}
\title{Novel Fermi Liquid of 2D Polar Molecules}

\author{Zhen-Kai Lu$^{1,2,3}$ and G. V. Shlyapnikov$^{2,4}$}
\affiliation{
\mbox{$^{1}$Max-Planck-Institut f\"ur Quantenoptik, Hans-Kopfermann-Stra{\ss}e 1, 85748 Garching, Germany}\\ 
\mbox{$^{2}$Laboratoire de Physique Th\'eorique et Mod\`eles Statistiques, CNRS and Universit\'e Paris Sud, UMR8626, 91405 Orsay, France}\\
\mbox{$^{3}$ D\'epartement de Physique, \'Ecole Normale Sup\'erieure, 75005, Paris, France}\\
\mbox{$^{4}$Van der Waals-Zeeman Institute, University of Amsterdam, Science Park 904, 1098 XH Amsterdam, The Netherlands}}
\begin{abstract}

We study Fermi liquid properties of a weakly interacting 2D gas of single-component fermionic polar molecules with dipole moments $d$ oriented perpendicularly to the plane of their translational motion. This geometry allows the minimization of inelastic losses due to chemical reactions for reactive molecules and, at the same time, provides a possibility of a clear description of many-body (beyond mean field) effects. The long-range character of the dipole-dipole repulsive interaction between the molecules, which scales as $1/r^3$ at large distances $r$, makes the problem drastically different from the well-known problem of the two-species Fermi gas with repulsive contact interspecies interaction. We solve the low-energy scattering problem and develop a many-body perturbation theory beyond the mean field. The theory relies on the presence of a small parameter $k_Fr_*$, where $k_F$ is the Fermi momentum, and $r_*=md^2/\hbar^2$ is the dipole-dipole length, with $m$ being the molecule mass. We obtain thermodynamic quantities as a series of expansion up to the second order in $k_Fr_*$ and argue that many-body corrections to the ground-state energy can be identified in experiments with ultracold molecules, like it has been recently done for ultracold fermionic atoms. Moreover, we show that only many-body effects provide the existence of zero sound and calculate the sound velocity. 

\end{abstract}
\date{\today}
\pacs{}
\maketitle

\section{Introduction}

The recent breakthrough in creating ultracold diatomic polar molecules in the ground ro-vibrational state \cite{Ni,Deiglmayr2008,Inouye,Nagerl} and cooling them towards quantum degeneracy \cite{Ni} has opened fascinating prospects for the observation of novel quantum phases 
\cite{Baranov2008,Lahaye2009,Pupillo2008,Wang2006,Buchler2007,Taylor,Cooper2009,Pikovski2010,Lutchyn,Potter,Barbara,Sun,Miyakawa,Gora,Ronen,Parish,Babadi,Baranov}. A serious problem in this direction is related to ultracold chemical reactions, such as  KRb+KRb$\Rightarrow$K$_2$+Rb$_2$ observed in the JILA experiments with KRb molecules \cite{Jin,Jin2}, which places severe limitations on the achievable density in three-dimensional samples. In order to suppress chemical reactions and perform evaporative cooling, it has been proposed to induce a strong dipole-dipole repulsion between the molecules by confining them to a (quasi)two-dimensional (2D) geometry and orienting their dipole moments (by a strong electric field) perpendicularly to the plane of the 2D translational motion \cite{Bohn1, Baranov1}. The suppression of chemical reactions by nearly two orders of magnitude in the quasi2D geometry has been demonstrated in the recent JILA experiment \cite{Ye}. At the same time, not all polar molecules of alkali atoms, on which experimental efforts are presently focused, may undergo these chemical reactions \cite{Jeremy}. In particular, they are energetically unfavorable for RbCs bosonic molecules obtained in Innsbruck \cite{Nagerl}, or for NaK and KCs molecules which are now being actively studied by several experimental groups (see, e.g. \cite{MITZ}). It is thus expected that future experimental studies of many-body physics will deal with non-reactive polar molecules or with molecules strongly confined to the 2D regime.

Therefore, the 2D system of fermionic polar molecules attracts a great deal of interest, in particular when they are in the same internal state. Various aspects have been discussed regarding this system in literature, in particular the emergence and beyond mean field description of the topological $p_x+ip_y$ phase for microwave-dressed polar molecules \cite{Cooper2009,Gora}, interlayer superfluids in bilayer and multilayer systems \cite{Pikovski2010,Ronen,Potter,Zinner}, the emergence of density-wave phases for tilted dipoles \cite{Sun,Miyakawa,Parish,Babadi,Baranov}. The case of superfluid pairing for tilted dipoles in the quasi2D geometry beyond the simple BCS approach has been discussed in Ref.~\cite{Baranov}. The Fermi liquid behavior of this system has been addressed by using the Fourier transform of the dipole-dipole interaction potential \cite{Das1,Das2,Miyakawa,Taylor,Pu,Das3,Baranov} and then employing various types of mean field approaches, such as the Hartree-Fock approximation \cite{Miyakawa} or variational approaches \cite{Taylor, Pu}. It should be noted, however, that the short-range physics can become important for the interaction between such polar molecules, since in combination with the long-range behavior it introduces a peculiar momentum dependence of the scattering amplitude \cite{Gora}. 

On the other hand, there is a subtle question of many-body (beyond mean field) effects in the Fermi liquid behavior of 2D polar molecules, and it can be examined in ultracold molecule experiments. For the case of atomic fermions, a milestone in this direction is the recent result at ENS, where the experiment demonstrated the many-body correction to the ground state energy of a short-range interacting two-species fermionic dilute gas \cite{Salomon1,Salomon2}. This correction was originally calculated by Huang, Lee, and Yang \cite{Huang,Lee} by using a rather tedious procedure. Later, it was found by Abrikosov and Khalatnikov \cite{Abr} in an elegant way based on the Landau Fermi liquid theory \cite{Landau}. 
 
In this paper, we study a weakly interacting 2D gas of fermionic polar molecules which are all in the same internal state. It is assumed that each molecule has an average dipole moment $d$ which is perpendicular to the plane of the translational motion, so that the molecule-molecule interaction at large separations $r$ is 
\begin{equation}  \label{VO}
U(r)=\frac{d^2}{r^3}=\frac{\hbar^2 r_*}{mr^3}, 
\end{equation}
where $r_*=md^2/\hbar^2$ is the characteristic dipole-dipole distance, and $m$ is the molecule mass. The value of $d$ depends on the external electric field. At ultralow temperatures that are much smaller than the Fermi energy, characteristic momenta of particles are of the order of the Fermi momentum $k_F$, and the criterion of the weakly interacting regime is:
\begin{equation}     \label{weakint}
k_Fr_*\ll 1.
\end{equation}

As a consequence, the Fermi liquid properties of this system, such as the ground state energy, compressibility, effective mass, can be written as a series of expansion in the small parameter $k_Fr_*$. We obtain explicit expressions of these quantities up to the second order in $k_Fr_*$, which requires us to reveal the role of the short-range physics in the scattering properties and develop a theory beyond the mean field. Our analysis shows that only many-body (beyond mean field) effects  provide the existence of undamped zero sound in the collisionless regime.

The paper is organized as follows. In Section II we analyze the low-energy 2D scattering of the polar molecules due to the dipole-dipole interaction. We obtain the  scattering amplitude for all scattering channels with odd orbital angular momenta. The leading part of the amplitude comes from the so-called anomalous scattering, that is the scattering related to the interaction between particles at distances of the order of their de Broglie wavelength. This part of the amplitude corresponds to the first Born approximation and, due to the long-range $1/r^3$ character of the dipole-dipole interaction, it is proportional to the relative momentum $k$ of colliding particles for any orbital angular momentum $l$. We then take into account the second Born correction, which gives a contribution proportional to $k^2$. For the $p$-wave scattering channel it is necessary to include the short-range contribution, which together with the second Born correction leads to the term behaving as $k^2\ln{k}$. In Section III, after reviewing the Landau Fermi liquid theory for 2D systems, we specify two-body (mean field) and many-body (beyond mean field) contributions to the ground state energy for 2D fermionic polar molecules in the weakly interacting regime. We then calculate the interaction function of quasiparticles on the Fermi surface and, following the idea of Abrikosov-Khalatnikov \cite{Abr}, obtain the compressibility, ground state energy, and effective mass of quasiparticles. In Section IV we calculate the zero sound velocity and stress that the many-body contribution to the interaction function of quasiparticles is necessary for finding the undamped zero sound. We conclude in Section V, emphasizing that the 2D gas of fermionic polar molecules represents a novel Fermi liquid, which is promising for revealing many-body effects. Moreover, we show that with present facilities it is feasible to obtain this system in both collisionless and hydrodynamic regimes. 

\section{Low-energy scattering of fermionic polar molecules in 2D}

\subsection{General relations}

We first discuss low-energy two-body scattering of identical fermionic polar molecules undergoing the 2D translational motion and interacting with each other at large separations via the potential $U(r)$ (\ref{VO}). The term low-energy means that their momenta satisfy the inequality $kr_*\ll 1$. In order to develop many-body theory for a weakly interacting gas of such molecules, we need to know the off-shell scattering amplitude defined as
\begin{equation}  \label{offshellgen}
f(\vec{k}^{\prime},\vec{k})=\int \exp(-i\vec{k}^{\prime}\vec{r})U(r)\tilde{\psi}_{\vec{k}}(\vec{r})d^{2}\vec{r},
\end{equation}
where $\tilde{\psi}_{\vec{k}}(\vec{r})$ is the true wavefunction of the relative motion with momentum $\vec{k}$. It is governed by the Schr\"odinger equation
\begin{equation}   \label{Schrgen}
\left(-\frac{\hbar^{2}}{m}\Delta+U(r)\right)\tilde{\psi}_{\vec{k}}(\vec{r})=\frac{\hbar^{2}k^2}{m}\tilde{\psi}_{\vec{k}}(\vec{r}).
\end{equation}
For $|{\bf k}'|=|{\bf k}|$ we have the on-shell amplitude which enters an asymptotic expression for $\psi_{\vec{k}}(\vec{r})$ at $r\rightarrow\infty$ \cite{Lan2,Gora}:
\begin{equation}    \label{psiassgen}
\tilde{\psi}_{\vec{k}}(\vec{r})=\exp(i{\bf kr})-\frac{m}{\hbar^2}\sqrt{\frac{i}{8\pi kr}}f(k,\varphi)\exp(ikr),
\end{equation}
with $\varphi$ being the scattering angle, i.e. the angle between the vectors ${\bf k}'$ and ${\bf k}$.

The wavefunction $\tilde\psi_{{\bf k}}({\bf r})$ can be represented as a sum of partial waves $\tilde\psi_l(k,r)$ corresponding to the motion with a given value of the orbital angular momentum $l$:
\begin{equation}    \label{psil}
\tilde\psi_{{\bf k}}({\bf r})=\sum_{l=-\infty}^{\infty}\tilde\psi_l(k,r)i^{l}\exp(il\varphi).
\end{equation}
Using the relation 
\begin{equation}    \label{expl}
\exp(i{\bf kr})=\sum_{l=-\infty}^{\infty} i^{l} J_{l}(kr)\exp[il(\varphi_{k}-\varphi_{r})],
\end{equation}
where $J_l$ is the Bessel function, and $\varphi_k$ and $\varphi_r$ are the angles of the vectors ${\bf k}$ and ${\bf r}$ with respect to the quantization axis. Eqs.~(\ref{psil}) and (\ref{expl}) allow one to express the scattering amplitude as a sum of partial-wave contributions: 
\begin{equation}    \label{offshellexp}
f(\vec{k}^{\prime},\vec{k})=\sum_{l=-\infty}^{\infty}\exp(il\varphi)f_{l}(k',k),
\end{equation}
with the off-shell $l$-wave amplitude given by
\begin{equation}     \label{offshelll}
f_l(k',k)=\int_0^{\infty}J_l(k'r)U(r)\tilde\psi_l(k,r)2\pi rdr.
\end{equation}
Similar relations can be written for the on-shell scattering amplitude:
\begin{eqnarray}   
&&f(k,\varphi)=\sum_{l=-\infty}^{\infty}\exp(il\varphi)f_{l}(k),  \label{fsum} \\
&&f_l(k)=\int_{0}^{\infty}J_{l}(k'r)U(r)\tilde{\psi}_{l}(k,r)2\pi rdr.  \label{flgen} 
\end{eqnarray}

The asymptotic form of the wavefunction of the $l$-wave relative motion at $r\rightarrow\infty$ may be represented as
\begin{equation}   \label{psideltal}
\tilde\psi_l(k,r)\propto\frac{\cos(kr-\pi/4+\delta_l(k))}{\sqrt{kr}},
\end{equation}
where $\delta_l(k)$ is the scattering phase shift. This is obvious because in the absence of scattering the $l$-wave part of the plane wave $\exp(i{\bf kr})$ at $r\rightarrow\infty$ is $(kr)^{-1/2}\cos(kr-\pi/4)$. Comparing Eq.~(\ref{psideltal}) with the $l$-wave part of Eq.~(\ref{psiassgen}) we obtain a relation between the partial on-shell amplitude and the phase shift:
\begin{equation}   \label{fldeltal}
f_{l}(k)=-\frac{4\hbar^{2}}{m}\frac{\tan \delta_l(k)}{1-i \tan \delta_l(k)}.
\end{equation}
Note that away from resonances the scattering phase shift is small in the low-momentum limit $kr_*\ll 1$. 

For the solution of the scattering problem it is more convenient to normalize the wavefunction of the radial relative motion with orbital angular momentum $l$ in such a way that it is real and for $r\rightarrow\infty$ one has:
\begin{eqnarray}
\label{nor}
\psi_l(k,r)&&=\left[J_l(kr)-\tan \delta_l(k)N_l(kr)\right] \nonumber \\
&&\propto \cos(kr-l\pi/2-\pi/4+\delta_l(k)),
\end{eqnarray}
where $N_l$ is the Neumann function. One checks straightforwardly that
$$\tilde\psi_l(k,r)=\frac{\psi_l(k,r)}{1-i\tan\delta_l(k)}.$$
Using this relation the off-shell scattering amplitude (\ref{offshelll}) can be represented as
\begin{equation}   \label{barfloff}
f_l(k',k)=\frac{{\bar f}_l(k',k)}{1-i\tan\delta_l(k)},
\end{equation}
where ${\bar f}_l(k',k)$ is real and follows from Eq.~(\ref{offshelll}) with $\tilde\psi_l(k,r)$ replaced by $\psi_l(k,r)$. Setting $k'=k$ we then obtain the related on-shell scattering amplitude:
\begin{equation}       \label{barflon}
{\bar f}_l(k,k)\equiv{\bar f}_l(k)=-\frac{4\hbar^2}{m}\tan \delta_l (k).
\end{equation} 

\subsection{Low-energy $p$-wave scattering}

As we will see, the slow $1/r^3$ decay of the potential $U(r)$ at sufficiently large distances makes the scattering drastically different from that of short-range interacting atoms. For identical fermionic polar molecules, only the scattering with odd orbital angular momenta $l$ is possible. For finding the amplitude of the $p$-wave scattering in the ultracold limit, $kr_*\ll 1$, we employ the method developed in Ref.~\cite{Gora} and used there for the scattering potential containing an attractive $1/r^3$ dipole-dipole tail. We divide the range of distances into two parts: $r<r_0$ and $r>r_0$, where $r_0$ is in the interval $r_*\ll r_0\ll k^{-1}$. In region I where $r<r_0$, the $p$-wave relative motion of two particles is governed by the Schr\"odinger equation with zero kinetic energy:
\begin{equation}   \label{Schrp1}
-\frac{\hbar^2}{m}\left(\frac{d^2\psi _I}{d r^2}+\frac{1}{r}\frac{d \psi_I}{d r}-\frac{\psi_I}{r^2}\right)+U(r)\psi_I=0.
\end{equation}
At distances where the potential $U(r)$ already acquires the form (\ref{VO}), the solution of Eq.~(\ref{Schrp1}) can be expressed in terms of growing and decaying Bessel functions:
\begin{equation}
\label{short}
\psi_I(r)\propto \left[AI_2\left(2\sqrt{\frac{r_*}{r}}\right)+K_2\left(2\sqrt{\frac{r_*}{r}}\right)\right].
\end{equation}
The constant $A$ is determined by the behavior of $U(r)$ at short distances where Eq.~(\ref{VO}) is no longer valid. If the interaction potential $U(r)$ has the form (\ref{VO}) up to very short distances, then $A=0$, so that for 
$r\rightarrow 0$ equation (\ref{short}) gives an exponentially decaying wavefunction. 

It should be noted here that for the quasi2D regime obtained by a tight confinement of the translational motion in one direction, we can encounter the situation where $r_*\lesssim l_0$, with $l_0$ being the confinement length. However, we may always select $r_0\gg l_0$ if the condition $kl_0\ll 1$ is satisfied. Therefore, our results for the 2D $p$-wave scattering obtained below in this section remain applicable for the quasi2D regime. The character of the relative motion of particles at distances $r\lesssim l_0$ is only contained in the value of the coefficient $A$, and the extra requirement is the inequality $kl_0\ll 1$.   

At large distances, $r>r_0$, the relative motion is practically free and the potential $U(r)$ can be considered as perturbation. To zero order, the relative wavefunction is given by
\begin{equation} \label{psiII0}
\psi_{II}^{(0)}(r)=J_1(kr)-\tan \delta_I(k)N_1(kr),
\end{equation}
where the phase shift $\delta_I(k)$ is due to the interaction between particles in region I. Equalizing the logarithmic derivatives of $\psi_I(r)$ and $\psi_{II}^{(0)}$ at $r=r_0$ we obtain:
\begin{equation}  \label{deltaI}
\!\!\tan \delta_I\!=\!-\frac{\pi k^2r_0r_*}{8}\left[1\!+\!\frac{r_*}{r_0}\left(2C\!-\!\frac{1}{2}\!-\!2A\!+\!\ln\frac{r_*}{r_0}\right)\right],\!\!
\end{equation}
with $C=0.5772$ being the Euler constant.

We now include perturbatively  the contribution to the $p$-wave scattering phase shift from distance $r>r_0$. In this region, to first order in $U(r)$, the relative wavefunction is given by
\begin{equation}
\label{sec}
\!\psi^{(1)}_{II}(r)=\psi^{(0)}_{II}(r)\!-\!\int_{r_0}^{\infty}G(r,r')U(r')\psi^{(0)}_{II}(r')2\pi r' dr'\!,
\end{equation}
where the Green function for the free $p$-wave motion obeys the radial equation:
\begin{eqnarray}
-\frac{\hbar^2}{m}\left(\frac{d^2}{dr^2}+\frac{1}{r}\frac{d}{dr}-\frac{1}{r^2}+k^2 \right)G(r,r')=\frac{\delta(r-r')}{2\pi r}. \nonumber
\end{eqnarray}
For the normalization of the relative wavefunction chosen in Eq.~(\ref{nor}), we have:
\begin{equation}
G(r,r')=-\frac{m}{4\hbar^2}
\begin{cases}
\psi^{(0)}_{II}(r')N_1(kr), &r>r'\\
\\
\psi^{(0)}_{II}(r)N_1(kr'). &r<r'
\end{cases}
\end{equation}
Substituting this Green function into Eq.~(\ref{sec}) and taking the limit $r\rightarrow \infty$, for the first order contribution to the phase shift we have:
\begin{equation}  \label{f1leadingint}
\!\!\tan \delta_1^{(1)}(k)\!=\!\tan\delta_I(k)\!-\!\frac{m}{4\hbar^2}\int_{r_0}^{\infty}\!\![\psi^{(0)}_{II}(r)]^2U(r)2\pi rdr.
\end{equation}
Using Eqs.~(\ref{psiII0}) and (\ref{deltaI}) we then obtain:
\begin{equation}
\!\!\!\tan\delta_1^{(1)}(k)\!=\!-\frac{2kr_*}{3}\!-\!\frac{\pi k^2 r_*^2}{8}\left(\!\!-2A\!+\!2C\!+\!\ln\frac{r_*}{r_0}\!-\!\frac{3}{2}\!\right).\!
\end{equation}

To second order in $U(r)$, we have the relative wavefunction:
\begin{eqnarray} \label{secondwave}
\psi^{(2)}_{II}(r)&&=\psi^{(1)}_{II}(r)+\int_{r_0}^{\infty} G(r,r')U(r')2\pi r' dr' \nonumber\\
 &&\times\int_{r_0}^{\infty} G(r',r'')U(r'')\psi^{(0)}_{II}(r'')2\pi r'' dr''.
\end{eqnarray} 
Taking the limit $r\rightarrow\infty$ in this equation we see that including the second order contribution, the scattering phase shift becomes:
\begin{eqnarray}
\tan\delta_1(k)&&=\tan\delta^{(1)}(k)-\frac{m^2}{8\hbar^4}\int_{r_0}^{\infty}\psi^{(0)}_{II}(r)^{2}U(r)2\pi r dr\nonumber \\
&&\times\int_{r}^{\infty}N_1(kr')U(r')\psi^{(0)}_{II}(r')2\pi r' dr'.
\end{eqnarray}
As we are not interested in terms that are proportional to $k^3$ or higher powers of $k$, we may omit the term $\tan\delta_I(k)N_1(kr)$ in the expression for $\psi^{(0)}_{II}(r)$. Then the integration over $dr'$ leads to: 
\begin{eqnarray}
&&\tan\delta_1(k)=\tan\delta_1^{(1)}(k)-\frac{(\pi kr_*)^2}{2}\int_{kr_0}^{\infty} \frac{J^2_1(x)}{x^2}dx\nonumber \\
&&\times \Big[\frac{2}{3}x\left(N_0(x)J_2(x)-N_1(x)J_1(x)\right) \nonumber\\
&& \;\;\;\;-\frac{1}{2}N_0(x)J_1(x)+\frac{1}{6}N_1(x)J_2(x)-\frac{1}{\pi x}\Big].
\end{eqnarray}
For the first four terms in the square brackets, we may put the lower limit of integration equal to zero and use the following relations:
\begin{align*}
&\int_{0}^{\infty} J^3_1(x)N_1(x)\frac{dx}{x}=-\frac{1}{4\pi},\\
&\int_{0}^{\infty} J^2_1(x)J_2(x)N_0(x)\frac{dx}{x}=\frac{1}{8\pi},\\
&\int_{0}^{\infty} J^3_1(x)N_0(x)\frac{dx}{x^2}=\frac{1}{16\pi},\\
&\int_{0}^{\infty} J^2_1(x)J_2(x)N_1(x)\frac{dx}{x^2}=-\frac{1}{16\pi}.
\end{align*}
For the last term in the square brackets we have:
\begin{equation}
\int_{kr_0}^{\infty} J^{2}_1(x)\frac{dx}{x^3}\approx\frac{1}{16}-\frac{C}{4}+\frac{\ln 2}{4}-\frac{1}{4}\ln kr_0.
\end{equation}
We then obtain:
\begin{align}
\tan\delta_1(k)&=\tan\delta_1^{(1)}(k)-\frac{\pi(kr_*)^2}{8}\left[\frac{7}{12}+C-\ln 2+\ln kr_0\right] \nonumber\\
&=-\frac{2kr_*}{3}-\frac{\pi k^2 r^2_*}{8}\ln\xi kr_*,  \label{deltap}
\end{align}
where:
\begin{equation}   \label{xi}
\xi=\exp\left(3C-\ln 2 -\frac{11}{12}-2A\right).
\end{equation}

Using Eqs.~(\ref{barflon}) and (\ref{deltap}) we represent the on-shell $p$-wave scattering amplitude ${\bar f}_1(k)$ in the form:
 \begin{equation}
{\bar f}_1(k)={\bar f}^{(1)}_1(k)+{\bar f}^{(2)}_1(k),
 \end{equation}
with
\begin{equation}    \label{f1leading}
{\bar f}^{(1)}_1(k)=\frac{8\hbar^2}{3m}kr_*
\end{equation}
and
\begin{equation} \label{f12}
{\bar f}^{(2)}_1(k)=\frac{\pi\hbar^2}{2m}(kr_*)^2\ln\xi kr_*.
\end{equation}
The leading term is ${\bar f}^{(1)}_1(k)\propto k$. It appears to first order in $U(r)$ and comes from the scattering at distances $r\sim 1/k$. This term can be called ``anomalous scattering'' term (see \cite{Lan2}). The term $f^{(2)}_1(k)\propto k^2\ln\xi kr_*$ comes from both large distances $\sim 1/k$ and short distances. Note that the behavior of the wavefunction at short distances where $U(r)$ is no longer given by Eq.~(\ref{VO}), is contained in Eq.~(\ref{deltap}) only through the coefficient $\xi$ under logarithm.   

\subsection{Scattering with $|l|>1$}

The presence of strong anomalous $p$-wave scattering, i.e. the scattering from interparticle distances $\sim 1/k$, originates from the slow $1/r^3$ decay of the potential $U(r)$ at large $r$. The strong anomalous scattering is also present for partial waves with higher $l$. In this section we follow the same method as in the case of the $p$-wave scattering and calculate the amplitude of the $l$-wave scattering with $|l|>1$. For simplicity we consider positive $l$, having in mind that the scattering amplitude and phase shift depend only on $|l|$.  

To zero order in $U(r)$, the wavefunction of the $l$-wave relative motion at large distances $r>r_0$ is written as:
\begin{equation}
\label{large}
\psi^{(0)}_{l(II)}(k,r)=\left[J_l(kr)-\tan \delta_{l(I)}(k)N_l(kr)\right],
\end{equation} 
where $\delta_{l(I)}(k)$ is the $l$-wave scattering phase shift coming from the interaction at distances $r<r_0$. We then match $\psi^{(0)}_{l(II)}(k,r)$ at $r=r_0$ with the short-distance wavefunction $\psi_{l(I)}(r)$ which follows from the Schr\"odinger equation for the $l$-wave relative motion in the potential $U(r)$ at $k=0$. This immediately gives a relation:
\begin{equation} \label{arg}
\tan \delta_{l(I)}(k)=\frac{k J'_l(kr_0)-w_lJ_l(kr_0)}{k N'_l(kr_0)-w_lN_l(kr_0)},
\end{equation}
where the momentum-independent quantity $w_l$ is the logarithmic derivative of $\psi_{l(I)}(r)$ at $r=r_0$. Since we have the inequality $kr_0\ll 1$, the arguments of the Bessel functions in Eq.~(\ref{arg}) are small 
and they reduce to $J_l(x)\sim x^l \; ,\; N_l(x)\sim x^{-l}$. This leads to $\tan \delta_{l(I)}(k) \sim (kr_0)^{2l}$. Thus, the phase shift coming from the interaction at short distances is of the order of $(kr_0)^{2l}$. As we confine ourselves to second order in $k$, we may put $\tan \delta_{l(I)}(k)=0$ for the scattering with $|l|>1$.

Then, like for the $p$-wave scattering, we calculate the contribution to the phase shift from distances $r>r_0$ by considering the potential $U(r)$ as perturbation. To first and second order in $U(r)$, at $r>r_0$ we have similar expressions as Eq. (\ref{f1leadingint}), (\ref{secondwave}) for the relative wavefunction of the $l$-wave motion.
Following the same method as in the case of the $p$-wave scattering and retaining only the terms up to $k^2$, for the first order phase shift we have:
\begin{eqnarray}    \label{flleadingint}
&&\hspace{-17mm}\tan \delta_l^{(1)}(k)=-\frac{m}{4\hbar^2}\int_{r_0}^{\infty}[\psi^{(0)}_{l(II)}(r)]^2U(r)2\pi rdr \nonumber\\
&&\simeq -\frac{\pi kr_*}{2}\int_{kr_0}^{\infty}J_l^2(x)\frac{1}{x^2} dx=-\frac{2 kr_*}{4 l^2-1}.
\end{eqnarray}
The second order phase shift is:
\begin{align}  \label{ex}
&&\tan\delta^{(2)}_l(k)=-\frac{m^2}{8\hbar^4}\int_{r_0}^{\infty}\psi^{(0)}_{l(II)}(r)^{2}U(r)2\pi r dr \nonumber \\
&&\;\;\times\int_{r}^{\infty}N_l(kr')U(r')\psi^{(0)}_{l(II)}(r')2\pi r' dr' \nonumber\\
&&\simeq -\frac{(\pi k r_*)^2}{2}\int_{kr_0}^{\infty}\frac{J^2_l(x)}{x^2} dx \int_{x}^{\infty}\frac{N_l(y)J_l(y)}{y^2} dy,
\end{align}
and we may put the lower limit of integration equal to zero.
For the integral over $dy$, we obtain :
\begin{align}
&\int_{x}^{\infty}\frac{N_l(y)J_l(y)}{y^2} dy \nonumber \\
&=\frac{1}{2l(2l-1)}J_{l}(x)N_{l-1}(x)+\frac{1}{2l(2l+1)}J_{l+1}(x)N_{l}(x) \nonumber\\
&\!\!+\frac{2x}{4l^2-1}\big[N_{l-1}(x)J_{l+1}(x)\!-\!J_l(x)N_l(x)\big]-\frac{1}{\pi lx}.
\end{align}
Then, using the relations:
\begin{equation*}
\int_{0}^{\infty} \frac{J_{l}^{2}(x)}{x^3} dx=\frac{1}{4l (l^2-1)},
\end{equation*}
\begin{equation*}
\int_{0}^{\infty}\frac{J_{l}^{2}(x)}{x} N_{l-1}(x) J_{l+1}(x) dx =\frac{1}{4l (l+1)\pi},
\end{equation*}
\begin{equation*}
\int_{0}^{\infty}\frac{J_{l}^{3}(x)}{x} N_{l}(x) dx =-\frac{1}{4l^2\pi},
\end{equation*}
\begin{equation*}
\int_{0}^{\infty}\frac{J_{l}^{2}(x)}{x^2} J_{l}(x)N_{l-1}(x) dx =\frac{1}{8l^2(l+1)\pi},
\end{equation*}
\begin{equation*}
\int_{0}^{\infty}\frac{J_{l}^{2}(x)}{x^2} J_{l+1}(x)N_{l}(x) dx =-\frac{1}{8l^2(l+1)\pi},
\end{equation*}
we find the following result for the second order phase shift:
\begin{eqnarray}
\tan\delta^{(2)}_l(k)=\frac{3\pi(kr_*)^2}{8}\frac{1}{l(l^2-1)(4l^2-1)}.
\end{eqnarray}
So, the total phase shift is given by
\begin{align}
\tan \delta_l(k)&=\tan \delta_l^{(1)}(k)+\tan\delta_l^{(2)}(k)\nonumber\\
&=-\frac{2 kr_*}{4l^2-1}+\frac{3\pi(kr_*)^2}{8l(l^2-1)(4l^2-1)}.
\end{align}
Then, according to Eq.~(\ref{barflon}) the on-shell scattering amplitude ${\bar f}_l(k)$ is
\begin{equation}
{\bar f}_l(k)={\bar f}^{(1)}_l(k)+{\bar f}^{(2)}_l(k),
 \end{equation}
where
\begin{equation}   \label{flleading}
{\bar f}^{(1)}_l(k)=\frac{8\hbar^2 kr_*}{m}\frac{1}{4 l^2-1},
\end{equation}
\begin{equation} \label{fl2}
{\bar f}^{(2)}_l(k)=-\frac{3\pi\hbar^2}{2m}(kr_*)^2\frac{1}{|l|(l^2-1)(4l^2-1)}.
\end{equation}
Note that Eqs.~(\ref{flleading}) and (\ref{fl2}) do not contain short-range contributions as those are proportional to $k^{2|l|}$ and can be omitted for $|l|>1$. 

\subsection{First order Born approximation and the leading part of the scattering amplitude}

As we already said above, in the low-momentum limit for both $|l|=1$ and $|l|>1$ the leading part of the on-shell scattering amplitude ${\bar f}_l(k)$ is ${\bar f}_l^{(1)}(k)$ and it is contained in the first order contribution from distances $r>r_0$. For $|l|>1$ it is given by Eq.~(\ref{flleading}) and follows from Eq.~(\ref{flleadingint}) with $\psi^{(0)}_{l(II)}=J_l(kr)$. In the case of $|l|=1$ this leading part is given by Eq.~(\ref{f1leading}) and follows from the integral term of Eq.~(\ref{f1leadingint}) in which one keeps only $J_1(kr)$ in the expression for $\psi^{(0)}_{II}(r)$. This means that ${\bar f}_l^{(1)}(k)$ actually follows from the first order Born approximation. 

The off-shell scattering amplitude can also be represented as ${\bar f}_l(k',k)={\bar f}_l^{(1)}(k',k)+{\bar f}_l^{(2)}(k',k)$, and the leading contribution ${\bar f}_l^{(1)}(k',k)$ follows from the first Born approximation. It is given by Eq.~(\ref{offshelll}) in which one should replace $\tilde\psi_l(k,r)$ by $J_l(kr)$:
\begin{equation}  \label{leadingoffshelll}
{\bar f}_l^{(1)}(k',k)=\int_{0}^{\infty}  J_{l}(kr) J_{l}(kr')U(r) 2\pi r dr.
\end{equation}
Note that it is not important that we put zero for the lower limit of the integration, since this can only give a correction which behaves as $k^2$ or a higher power of $k$. Then, putting $U(r)=\hbar^2r_*/mr^3$ in Eq.~(\ref{leadingoffshelll}), we obtain: 
\begin{align}
\label{off}
{\bar f}^{(1)}_l(k',k)= &\frac{\pi \hbar^2}{m}\frac{\Gamma(l-1/2)}{\sqrt{\pi}}\frac{k^l r_*}{(k')^{l-1}} \nonumber \\
& \times F\left(-\frac{1}{2},-\frac{1}{2}+l,1+l,\frac{k^2}{k'^2}\right),
\end{align}
where $F$ is the hypergeometric function. The result of Eq.~(\ref{off}) corresponds to $k<k'$, and for $k>k'$ one should interchange $k$ and $k'$. 

For identical fermions the full scattering amplitude contains only partial amplitudes with odd $l$. Since the scattered waves with relative momenta ${\bf k}'$ and $-{\bf k}'$ correspond to interchanging the identical fermions, the scattering amplitude can be written as (see, e.g. \cite{Lan2}):
\begin{equation}     \label{ffermion}
\tilde f({\bf k}',{\bf k})=f({\bf k}',{\bf k})-f(-{\bf k}',{\bf k}).
\end{equation}
Then, according to equation (\ref{fsum}) one can write:
\begin{equation}      \label{ffermionsum}
\tilde f({\bf k}',{\bf k})=2\sum_{l\,odd}f_l(k',k)\exp(il\varphi).
\end{equation}

In the first Born approximation there is no difference between $f_l(k',k)$ and ${\bar f}_l(k',k)$ because $\tan\delta_l(k)$ in the denominator of Eq.~(\ref{barfloff}) is proportional to $k$ and can be disregarded. Therefore, one may use ${\bar f}_l^{(1)}(k',k)$ of Eq.(\ref{off}) for $f_l(k',k)$ in Eq.~(\ref{ffermionsum}). One can represent $\tilde f({\bf k}',{\bf k})$ in a different form recalling that in the first Born approximation we have:
\begin{equation}       \label{fBorn}
f({\bf k}',{\bf k})=\int U(r)\exp[i({\bf k}-{\bf k}'){\bf r}]d^2r.
\end{equation}
Performing the integration in this equation, with $U(r)$ given by Eq.~(\ref{VO}), and using Eq.~(\ref{ffermion}) we obtain:
\begin{equation}    \label{tildefBorn}
\tilde f({\bf k}',{\bf k})=\frac{2\pi\hbar^2r_*}{m}\{|{\bf k}+{\bf k}'|-|{\bf k}-{\bf k}'|\}.
\end{equation}
Equation (\ref{tildefBorn}) is also obtained by a direct summation over odd $l$ in Eq.~(\ref{ffermionsum}), with $f_l(k',k)$ following from Eq.~(\ref{off}).

\section{Thermodynamics of a weakly interacting 2D gas of fermionic polar molecules at $T=0$}
\subsection{General relations of Fermi liquid theory}

Identical fermionic polar molecules undergoing a two-dimensional translational motion and repulsively interacting with each other via the potential (\ref{VO}) represent a 2D Fermi liquid. General relations of the Landau Fermi liquid theory remain similar to those in 3D (see, e.g. \cite{Landau}). The number of ``dressed" particles, or quasiparticles, is the same as the total number of particles $N$, and the (quasi)particle Fermi momentum is
\begin{equation}        \label{kF}
k_F=\sqrt{\frac{4\pi N}{S}},
\end{equation}
where $S$ is the surface area. At $T=0$ the momentum distribution of free quasiparticles is the step function
\begin{equation}   \label{nstep}
n({\bf k})=\theta(k_F-k),
\end{equation}
i.e. $n({\bf k})=1$ for $k<k_F$ and zero otherwise.The chemical potential is equal to the boundary energy at the Fermi circle, $\mu=\epsilon_F\equiv\epsilon(k_F)$.

The quasiparticle energy $\epsilon({\bf k})$ is a variational derivative of the total energy with respect to the distribution function $n({\bf k})$. Due to the interaction between quasiparticles, the deviation $\delta n$ of this distribution from the step function (\ref{nstep}) results in the change of the quasiparticle energy:
\begin{equation}
\label{1}
\delta\epsilon (\vec{k})=\int F(\vec{k},\vec{k}')  \delta n(\vec{k}') \frac{d^2k'}{(2\pi)^2}.
\end{equation}
The interaction function of quasiparticles $F(\vec{k},\vec{k}')$ is thus the second variational derivative of the total energy with regard to $n({\bf k})$. The quantity $\delta n({\bf k})$ is significantly different from zero only near the Fermi surface, so that one may put ${\bf k}=k_F{\bf n}$ and ${\bf k}'=k_F{\bf n}'$ in the arguments of $F$ in Eq.~(\ref{1}), where ${\bf n}$ and ${\bf n}'$ are unit vectors in the directions of ${\bf k}$ and ${\bf k}'$. 
The quasiparticle energy near the Fermi surface can be written as: 
\begin{equation} \label{2}
\epsilon(\vec{k})=\epsilon_{F}+\hbar v_{F}(k-k_{F})+\int F(\vec{k},\vec{k}')\delta n(\vec{k}')\frac{d^2k'}{(2\pi)^2}.
\end{equation} 
The quantity $v_F=\partial\epsilon({\bf k})/\hbar\partial k|_{k=k_F}$ is the Fermi velocity, and the effective mass of a quasiparticle is defined as $m^*=\hbar k_F/v_F$. It can be obtained from the relation (see \cite{Landau}):
\begin{align}
\label{eff}
\frac{1}{m}=\frac{1}{m^{*}}+\frac{1}{(2\pi\hbar)^{2}}\int_0^{2\pi} F(\theta) \cos\theta  d \theta,
\end{align}
where $\theta$ is the angle between the vectors ${\bf n}$ and ${\bf n}'$, and $F(\theta)=F(k_F{\bf n},k_F{\bf n}')$.

The compressibility $\kappa$ at $T=0$ is given by \cite{Landau}:
\begin{equation}   \label{kappa-1}
\kappa^{-1}=\frac{N^2}{S}\frac{\partial \mu}{\partial N}.
\end{equation}
The chemical potential is $\mu=\epsilon_{F}$, and the variation of $\mu$ due to a change in the number of particles can be expressed as
\begin{equation}
\label{mu}
\delta \mu=\int F(k_F{\bf n},\vec{k}')\delta n(\vec{k'})\frac{d^2k'}{(2\pi)^2}+\frac{\partial \epsilon_{F}}{\partial k_{F}} \delta k_{F}.
\end{equation}
The quantity $\delta n(\vec{k'})$ is appreciably different from zero only when $\vec{k'}$ is near the Fermi surface, so that we can replace the interaction function $F$ by its value on the Fermi surface. Then the first term of Eq.~(\ref{mu}) becomes
\begin{eqnarray}
\int F(\theta) \frac{d\theta}{2\pi}\int\delta n(\vec{k'})\frac{d^2k'}{(2\pi)^2}=\frac{\delta N}{2\pi S}\int F(\theta) d \theta. \nonumber
\end{eqnarray}
The second term of Eq.~(\ref{mu}) reduces to
\begin{eqnarray}
\frac{\partial \epsilon_{F}}{\partial k_{F}} \delta k_{F}=\frac{\hbar^2 k_{F}}{m^{*}}\delta k_{F}=\frac{2\pi\hbar^{2}}{m^{*}}\frac{\delta N}{S}.
\end{eqnarray}
We thus have (see \cite{Landau}):
\begin{align}
\label{u1}
\frac{\partial \mu}{\partial N}&=\frac{1}{2\pi S}\int_0^{2\pi} F(\theta) d\theta +\frac{2\pi \hbar^{2}}{m^{*}S} \nonumber\\
&=\frac{2\pi\hbar^{2}}{mS}+\frac{1}{2\pi S}\int_0^{2\pi} (1-\cos\theta) F(\theta) d\theta. 
\end{align}
Equation (\ref{u1}) shows that the knowledge of the interaction function of quasiparticles on the Fermi surface, $F(\theta)$, allows one to calculate $\partial\mu/\partial N$ and, hence, the chemical potential $\mu=\partial E/\partial N$ and the ground state energy $E$. This elegant way of finding the ground state energy has been proposed by Abrikosov and Khalatnikov \cite{Abr}. It was implemented in Ref.~\cite{Abr} for a two-component 3D Fermi gas with a weak repulsive contact (short-range) interspecies interaction.

We develop a theory beyond the mean field for calculating the interaction function of quasiparticles for a single-component 2D gas of fermionic polar molecules in the weakly interacting regime. We obtain the ground state energy as a series of expansion in the small parameter $k_Fr_*$ and confine ourselves to the second order. In this sense our work represents a sort of Lee-Huang-Yang \cite{Huang,Lee} and Abrikosov-Khalatnikov \cite{Abr} calculation for this dipolar system. As we will see, the long-range character of the dipole-dipole interaction makes the result quite different from that in the case of short-range interactions. 

\subsection{Two-body and many-body contributions to the ground state energy}

We first write down the expression for the kinetic energy and specify two-body (mean field) and many-body (beyond mean field) contributions to the interaction energy. The Hamiltonian of the system reads:
\begin{equation}
\label{ha}
\!\!\hat{\cal H}\!=\!\sum_{\vec{k}}\frac{\hbar^2k^2}{2m}\hat{a}_{\vec{k}}^{\dag}\hat{a}_{\vec{k}}\!+\!\frac{1}{2S}\!\!\!\sum_{\vec{k_{1}},\vec{k_{2}},\vec{q}}\!\!\!U(\vec{q}) \hat{a}_{\vec{k_{1}}+\vec{q}}^{\dag}\hat{a}_{\vec{k_{2}}-\vec{q}}^{\dag}\hat{a}_{\vec{k_{2}}}\hat{a}_{\vec{k_{1}}},\!\!
\end{equation}
where $\hat {a}^{\dagger}_{{\bf k}}$ and $\hat{ a}_{{\bf k}}$ are creation and annihilation operators of fermionic polar molecules, and $U(\vec{q})$ is the Fourier transform of the interaction potential $U(r)$: 
\begin{equation}   \label{Uq}
U(\vec q)=\int d^2\vec{r} U(r)e^{-i\vec{q}\cdot\vec{r}},
\end{equation}
The first term of Eq.~(\ref{ha}) represents the kinetic energy and it gives the main contribution to the total energy $E$ of the system. This term has only diagonal matrix elements, and using the momentum distribution (\ref{nstep}) at $T=0$ we have:
\begin{align}
\label{Ekin}
\frac{E_{kin}}{S}=\int_0^{k_F}\frac{\hbar^2k^2}{2m}\frac{2\pi kdk}{(2\pi)^2}=\frac{\hbar^2k_F^4}{16m}.
\end{align} 

The interaction between the fermionic molecules is described by the second term in Eq.~(\ref{ha}) and compared to the kinetic energy it provides a correction to the total energy $E$. The first order correction is given by the diagonal matrix element of the interaction term of the Hamiltonian:
\begin{align}
 \label{E1}
E^{(1)}&=\frac{1}{2S}\sum_{{\bf{k_1}},{\bf{k_2} },{\bf q}}U(\vec{q})\langle \hat{a}_{\vec{k_{1}}+\vec{q}}^{\dag}\hat{a}_{\vec{k_{2}}-\vec{q}}^{\dag}\hat{a}_{\vec{k_{2}}}\hat{a}_{\vec{k_{1}}}\rangle \nonumber\\
&=\frac{1}{2S}\sum_{\vec{k_{1}},\vec{k_{2}}}\left[U(0)-U(\vec{k_{2}}-\vec{k_{1}})\right]n_{\vec{k_{1}}}n_{\vec{k_{2}}}.
\end{align}
The second order correction to the energy of the state $\ket{j}$ of a non-interacting system can be expressed as:
\begin{eqnarray}
E_{j}^{(2)}=\sum_{m\neq j} \frac{V_{jm}V_{mj}}{E_{j}-E_{m}},
\end{eqnarray}
where the summation is over eigenstates $\ket{m}$ of the non-interacting system, and $V_{jm}$ is the non-diagonal matrix element. In our case the symbol $j$ corresponds to the ground state and the symbol $m$ to excited states. The non-diagonal matrix element is 
\begin{equation}
\!\!V_{jm}\!=\!\frac{1}{2S}\left<\!m\left|\sum_{\vec{k_{1}},\vec{k_{2}},\vec{q}} U(\vec{q}) \hat{a}_{\vec{k_{1}}+\vec{q}}^{\dag}\hat{a}_{\vec{k_{2}}-\vec{q}}^{\dag}\hat{a}_{\vec{k_{2}}}\hat{a}_{\vec{k_{1}}}\right|j\!\right>. \!
\end{equation}
This matrix element corresponds to the scattering of two particles from the initial state $\vec{k_{1}}$, $\vec{k_{2}}$ to an intermediate state $\vec{k'_{1}}$, $\vec{k'_{2}}$, and the matrix element $V_{mj}$ describes the reversed process in which the two particles return from the intermediate to initial state. Taking into account the momentum conservation law ${\bf k}_1+{\bf k}_2={\bf k}'_1+{\bf k}'_2$ the quantity $V_{jm}V_{mj}=|V_{jm}|^2$ is given by
\begin{align}
|V_{jm}|^2&=\frac{1}{(2S)^2}n_\vec{k_{1}}n_\vec{k_{2}}(1-n_\vec{k_{1}'})(1-n_\vec{k_{2}'})\nonumber\\
&\times\left|U(\vec{k'_{1}}-\vec{k_{1}})-U(\vec{k'_{2}}-\vec{k_{1}})\right|^2,
\end{align}
and the second order correction to the ground state energy takes the form:
\begin{align}
\label{E2}
E^{(2)}=&\frac{1}{(2S)^2}\sum_{\vec{k_{1}},\vec{k_{2}},\vec{k_{1}'}} \Bigg[\left| U({\bf k}'_1-{\bf k}_1)-U({\bf k}'_2-{\bf k}_1)\right|^{2} \nonumber\\
&\times \frac{n_\vec{k_{1}}n_\vec{k_{2}}(1-n_\vec{k_{1}'})(1-n_\vec{k_{2}'})}{\hbar^2(\vec{k^{2}_{1}}+\vec{k^{2}_{2}}-\vec{k'^{2}_{1}}-\vec{k'^{2}_{2}})/2m}\Bigg].
\end{align}

From Eq.~(\ref{E2}) we see that the second order correction diverges because of the term proportional to $n_\vec{k_{1}}n_\vec{k_{2}}$, which is divergent at large $k'_1$. This artificial divergence is eliminated by expressing the energy correction in terms of a real physical quantity, the scattering amplitude. The relation between the Fourier component of the interaction potential and the off-shell scattering amplitude is given by \cite{Lan2}:
\begin{equation}   \label{renorm}
f(\vec{k'},\vec{k})=U(\vec{k}'-\vec{k})+\frac{1}{S}\sum_{\vec{k''}}\frac{U(\vec{k}'-\vec{k''})f(\vec{k''},\vec{k})}{(E_{\vec{k}}-E_{\vec{k''}}-i0)},
\end{equation}  
where $E_\vec{k}=\hbar^2{\bf k}^2/m$ and $E_\vec{k''}=\hbar^2{\bf k}''^2/m$ are relative collision energies. Obviously, we have: $E_{\vec{k}}-E_{\vec{k''}}=\hbar^2(\vec{k_1}^2+\vec{k_2}^2-\vec{k''_1}^2-\vec{k''_2}^2)/2m $, with ${\bf k}_1$, ${\bf k}_2$ (${\bf k}''_1$, ${\bf k}''_2$) being the momenta of colliding particles in the initial (intermediate) state, as the relative momenta are given by $\vec{k}=(\vec{k}_1-\vec{k}_2)/2$, $\vec{k''}=(\vec{k''_1}-\vec{k''_2})/2$. We thus can write:
\begin{equation}
\label{ren}
\!\!\!U(\vec{k}'\!-\vec{k})\!=\!f(\vec{k}'\!,\!\vec{k})\!-\!\frac{2m}{\hbar^2 S}\sum_{\vec{k_{1}''}}\frac{U(\vec{k}'\!-\!\vec{k}'')f(\vec{k''}\!,\!\vec{k})}{\vec{k^{2}_{1}}\!+\!\vec{k^{2}_{2}}\!-\!\vec{k''^{2}_{1}}\!-\!\vec{k''^{2}_{2}}\!-\!i0}.\!\!
\end{equation}

Then, putting ${\bf k}'={\bf k}$ we have
\begin{eqnarray} 
U(0)=f({\bf k},{\bf k})-\frac{2m}{\hbar^2S}\sum_{{\bf k}''_1}\frac{U({\bf k}-{\bf k}'')f({\bf k}'',{\bf k})}{\vec{k^{2}_{1}}+\vec{k^{2}_{2}}-\vec{k''^{2}_{1}}-\vec{k''^{2}_{2}}-i0}, \nonumber
\end{eqnarray}
and setting ${\bf k}'=-{\bf k}$ we obtain
\begin{eqnarray}
\!\!U({\bf k}_2\!-\!{\bf k}_1)=f(-{\bf k},{\bf k})\!-\!\frac{2m}{\hbar^2S}\sum_{{\bf k}''_1}\frac{U(-{\bf k}\!-\!{\bf k}'')f({\bf k}'',{\bf k})}{\vec{k^{2}_{1}}\!+\!\vec{k^{2}_{2}}\!-\!\vec{k''^{2}_{1}}\!-\!\vec{k''^{2}_{2}}\!-\!i0}, \nonumber
\end{eqnarray}
Using these relations the first order correction (\ref{E1}) takes the form:
\begin{align}
\label{1str}
&E^{(1)}=\frac{1}{2S}\sum_{\vec{k_{1}},\vec{k_{2}}}\left[f(\vec{k},\vec{k})-f(\vec{-k},\vec{k})\right]n_{\vec{k_{1}}}n_{\vec{k_{2}}} \nonumber \\
&\!\!-\frac{1}{2S^2}\!\!\!\!\!\sum_{\vec{k_{1}}, \vec{k_{2}},\vec{k'_{1}}}\!\!\frac{[U(\vec{k}\!-\!\vec{k'})\!-\!U({\!-\bf k}\!-\!{\bf k}')]f(\vec{k'},\vec{k})}{\hbar^2(\vec{k^{2}_{1}}\!+\!\vec{k^{2}_{2}}\!-\!\vec{k'^{2}_{1}}\!-\!\vec{k'^{2}_{2}}\!-\!i0)/2m}n_{{\bf k}_1}n_{{\bf k}_2}. \!\!
\end{align}

The quantity $[U({\bf k}-{\bf k}')-U(-{\bf k}-{\bf k}')]$ in the second term of Eq.~(\ref{1str}), being expanded in circular harmonics $\exp(il\varphi)$ contains terms with odd $l$. Therefore, partial amplitudes with even $l$ in the expansion of the multiple $f({\bf k}',{\bf k})$ vanish after the integration over $d^2k'$. Hence, this amplitude can be replaced by $[f({\bf k}',{\bf k})-f({\bf k}',-{\bf k})]/2$. As we are interested only in the terms that behave themselves as $\sim k$ or $\sim k^2$, the amplitudes in the second term of Eq.~(\ref{1str}) are the ones that follow from the first Born approximation and are proportional to $k$. Therefore, we may put  $[U({\bf k}-{\bf k}')-U(-{\bf k}-{\bf k}')]=[f({\bf k},{\bf k}')-f(-{\bf k},{\bf k}')]$ and $f({\bf k}',{\bf k})=f^*({\bf k},{\bf k}')$. Then the first order correction takes the form:
\begin{align}   \label{1str1}
&E^{(1)}=\frac{1}{2S}\sum_{\vec{k_{1}},\vec{k_{2}}}\left[f(\vec{k},\vec{k})-f(\vec{-k},\vec{k})\right]n_{\vec{k_{1}}}n_{\vec{k_{2}}}-\frac{1}{(2S)^2}\nonumber \\
&\times\!\!\sum_{\vec{k_{1}}, \vec{k_{2}},\vec{k'_{1}}}\frac{|f(\vec{k'},\vec{k})\!-\!f({\bf k}',-{\bf k})|^2}{\hbar^2(\vec{k^{2}_{1}}\!+\!\vec{k^{2}_{2}}\!-\!\vec{k'^{2}_{1}}\!-\!\vec{k'^{2}_{2}}\!-\!i0)/2m}n_{{\bf k}_1}n_{{\bf k}_2}.
\end{align}

Using the expansion of the full scattering amplitude in terms of partial amplitudes as given by Eq.~(\ref{ffermionsum}) we represent the first order correction as
\begin{align}    \label{E1l}
&E^{(1)}=\frac{1}{S}\sum_{{\bf k}_1,{\bf k}_2}\sum_{l\,odd}f_l(k)n_{{\bf k}_1}n_{{\bf k}_2}-\frac{1}{S^2}\sum_{{\bf k}_1,{\bf k}_2}\sum_{l\,odd}\nonumber\\
&\!\times\!\!\int\! \frac{d^2k'}{(2\pi)^2}\frac{f_l^2(k)}{\hbar^2(\vec{k^{2}_{1}}\!+\!\vec{k^{2}_{2}}\!-\!\vec{k'^{2}_{1}}\!-\!\vec{k'^{2}_{2}}\!-\!i0)/2m}n_{{\bf k}_1}n_{{\bf k}_2}.\!\!
\end{align}
The contribution of the pole in the integration over $d^2k'$ in the second term of Eq.~(\ref{E1l}) gives $imf_l^2(k)/4\hbar^2$ for each term in the sum over ${\bf k}_1$, ${\bf k}_2$, and $l$, and we may use here the amplitude ${\bar f}^{(1)}_l(k)$. In the first term of Eq.~(\ref{E1l}) we should use $f_l(k)=f_l^{(1)}(k)+f_l^{(2)}(k)$. However, we may replace $f_l^{(2)}$ by ${\bar f}_l^{(2)}$ because the account of $\tan\delta(k)$ in the denominator of Eq.~(\ref{barfloff}) leads to $k^3$ terms and terms containing higher powers of $k$. For the amplitude $f_l^{(1)}(k)$, we use the expression:
$$f_l^{(1)}(k)={\bar f}^{(1)}_l+i\tan\delta(k){\bar f}^{(1)}_l={\bar f}^{(1)}_l-im[{\bar f}^{(1)}_l]^2/4\hbar^2,$$ 
which assumes a small scattering phase shift. The second term of this expression, being substituted into the first line of Eq.~(\ref{E1l}), exactly cancels the contribution of the pole in the second term of (\ref{E1l}). Thus, we may use the amplitude ${\bar f}_l$ in the first term of equation (\ref{E1l}) and take the principal value of the integral in the second term. The resulting expression for the first order correction reads:
\begin{align}    \label{1str2}
E^{(1)}&=\frac{1}{S}\sum_{{\bf k}_1,{\bf k}_2}{\bar f}({\bf k})n_{{\bf k}_1}n_{{\bf k}_2}-\frac{1}{(2S)^2}\nonumber\\
&\times\sum_{\vec{k_{1}}, \vec{k_{2}},\vec{k'_{1}}} \frac{2m|f(\vec{k},\vec{k'})-f({-\bf k},{\bf k}')|^2}{\hbar^2(\vec{k^{2}_{1}}+\vec{k^{2}_{2}}-\vec{k'^{2}_{1}}-\vec{k'^{2}_{2}})}n_{{\bf k}_1}n_{{\bf k}_2},
\end{align}
where ${\bar f}({\bf k})=\sum_{l\,\,odd}{\bar f}_l(k)$.

The second order correction (\ref{E2}) can also be expressed in terms of the scattering amplitude by using Eq.(\ref{renorm}). Replacing $U({\bf k}_1-{\bf k}'_1)=U({\bf k}-{\bf k}')$ and $U({\bf k}'_2-{\bf k}_1)=U(-{\bf k}-{\bf k}')$ by
$f({\bf k}',{\bf k})$ and $f(-{\bf k},{\bf k}')$, respectively, we have:
\begin{align}
\label{2ndr}
E^{(2)}=\frac{1}{(2S)^2}&\sum_{\vec{k_{1}},\vec{k_{2}},\vec{k_{1}'}}\Big[\frac{ \left| f(\vec{k'},\vec{k})- f(\vec{k'},-\vec{k})\right|^{2}}{\hbar^2(\vec{k^{2}_{1}}+\vec{k^{2}_{2}}-\vec{k'^{2}_{1}}-\vec{k'^{2}_{2}})/2m}\nonumber \\
&\times n_\vec{k_{1}}n_\vec{k_{2}}(1-n_\vec{k_{1}'})(1-n_\vec{k_{2}'})\Big].
\end{align}
 
Note that the divergent term proportional to $n_{{\bf k}_1} n_{{\bf k}_2}$ in Eq.~(\ref{2ndr}) and the (divergent) second term of Eq.~(\ref{1str2}) exactly cancel each other, and the sum of the first and second order corrections can be represented as $E^{(1)}+E^{(2)}=\tilde E^{(1)}+\tilde E^{(2)},$
where
\begin{equation} \label{tildeE1}
\tilde E^{(1)}=\frac{1}{S}\sum_{{\bf k}_1,{\bf k}_2}{\bar f}({\bf k})n_{{\bf k}_1}n_{{\bf k}_2},
\end{equation}
and
\begin{align}    \label{tilde2}
\tilde E^{(2)}=\frac{1}{(2S)^2}&\sum_{\vec{k_{1}}, \vec{k_{2}},\vec{k'_{1}}}\Big\{\frac{|f(\vec{k'},\vec{k})-f({\bf k'},-{\bf k})|^2}{\hbar^2(\vec{k^{2}_{1}}+\vec{k^{2}_{2}}-\vec{k'^{2}_{1}}-\vec{k'^{2}_{2}})/2m} \nonumber\\
&\times n_{{\bf k}_1}n_{{\bf k}_2}[(1-n_{{\bf k}'_1})(1-n_{{\bf k}'_2})-1]\Big\}.
\end{align}
The term $\tilde E^{(1)}$ originates from the two-body contributions to the interaction energy and can be quoted as the mean field term. The term $\tilde E^{(2)}$ is the many-body contribution, which is beyond mean field.

It is worth noting that the term proportional to the product of four occupation numbers vanishes because its numerator is symmetrical and the denominator is antisymmetrical with respect to an interchange
of ${\bf k}_1,{\bf k}_2$ and ${\bf k}'_1,{\bf k}'_2$. The terms containing a product of three occupation numbers, $n_{{\bf k}_1}n_{{\bf k}_2}n_{{\bf k}'_1}$ and $n_{{\bf k}_1}n_{{\bf k}_2}n_{{\bf k}'_2}$ are equal to each other because the denominator is symmetrical with respect to an interchange of ${\bf k}'_1$ and ${\bf k}'_2$. We thus reduce Eq.~(\ref{tilde2}) to
\begin{equation}    \label{tildeE2}
\!\!\!\!\tilde E^{(2)}\!\!=\!\!-\frac{1}{2S^2}\!\!\!\!\sum_{\vec{k_{1}},\vec{k_{2}},\vec{k'_{1}}}\!\!\!\frac{2m|f(\vec{k'},\vec{k})\!-\!f({\bf k'},\!-{\bf k})|^2}{\hbar^2(\vec{k^{2}_{1}}\!+\!\vec{k^{2}_{2}}\!-\!\vec{k'^{2}_{1}}\!\!-\!\vec{k'^{2}_{2}})}n_{{\bf k}_1}\!n_{{\bf k}_2}\!n_{{\bf k}'_1}. \!\!\!\!
\end{equation}
Equations (\ref{tildeE1}) and (\ref{tildeE2}) allow a direct calculation of the ground state energy. With respect to the mean field term $\tilde E^{(1)}$ this is done in Appendix~\ref{App1}. However, a direct calculation of the many-body correction $\tilde E^{(2)}$ is even a more tedious task than in the case of two-component fermions with a contact interaction. We therefore turn to the Abrikosov-Khalatnikov idea of calculating the ground state energy (and other thermodynamic quantities) through the interaction function of quasiparticles on the Fermi surface.

\subsection{Interaction function of quasiparticles}

The interaction function of quasiparticles $F({\bf k},{\bf k}')$ is the second variational derivative of the total energy with respect to the distribution $n_{{\bf k}}$. The kinetic energy of our system is linear in $n_{{\bf k}}$ (see Eq.~(\ref{ha})), and the second variational derivative is related to the variation of the interaction energy $\tilde E$. We have \cite{Landau}: 
\begin{equation}  \label{tildeF}
\delta \tilde E=\frac{1}{2S}\sum_{{\bf k},{\bf k}'}F({\bf k},{\bf k}')\delta n_{\vec{k}} \delta n_{\vec{k'}},
\end{equation}
where $\tilde E=\tilde E^{(1)}+\tilde E^{(2)}$, and the quantities $\tilde E^{(1)}$ and $\tilde E^{(2)}$ are given by equations (\ref{tildeE1}) and (\ref{tildeE2}). On the Fermi surface we should put $|{\bf k}|=|{\bf k}'|=k_F$, so that the interaction function will depend only on the angle $\theta$ between ${\bf k}$ and $\vec{k'}$. Hereinafter it will be denoted as $\tilde F(\theta)$.  

The contribution $\tilde F^{(1)}(\theta)=2S\delta\tilde E^{(1)}/\delta n_{{\bf k}}\delta n_{{\bf k}'}$ is given by
\begin{equation}     \label{tildef1}
\tilde F^{(1)}(\theta)=2f\left(\frac{|{\bf k}-{\bf k}'|}{2}\right)=2\sum_{l\,odd}{\bar f}_l\left(k_F|\sin{\frac{\theta}{2}}|\right),
\end{equation}
where ${\bar f}_l={\bar f}_l^{(1)}+{\bar f}_l^{(2)}$, and the amplitudes ${\bar f}_l^{(1)}$ and ${\bar f}_l^{(2)}$ follow from Eqs.~(\ref{f1leading}) and (\ref{f12}) at $|l|=1$, and from 
Eqs.~(\ref{flleading}), (\ref{fl2}) at $|l|>1$. We thus may write equation (\ref{flleading}),
$${\bar f}^{(1)}_{l}(k)=\frac{8\hbar^2}{m}\frac{1}{4 l^2-1} kr_*,$$
for any odd $l$, and 
\begin{eqnarray}
 {\bar f}^{(2)}_l(k)=\frac{\pi\hbar^2}{2m}(kr_*)^2\times
	 \begin{cases}
  		   \ln(\xi kr_{*});  & \text{$|l|=1$}\\
		   -\frac{3}{|l|(l^2-1)(4l^2-1)};  & \text{$|l|>1$}
  	\end{cases}  \nonumber
 \end{eqnarray}
with $\xi$ from Eq.~(\ref{xi}). Making a summation over all odd $l$ we obtain:
\begin{equation}   
{\bar f}^{(1)}(k)=\sum_{l\,odd} f^{(1)}_{l}(k)=\frac{2\pi\hbar^2}{m}kr_*,  \label{barf1} \\
\end{equation}
\begin{equation}
\!\!\!{\bar f}^{(2)}(k)\!=\!\!\!\sum_{l\,odd} f^{(2)}_{l}(k)\!\!=\!\!\frac{\pi\hbar^2}{m}(kr_{*}\!)^2\!\!\left[\ln(\xi kr_{*}\!)\!-\!\frac{25}{12}\!+\!3\!\ln 2\!\right]\!\!.\!\!\!\label{barf2}
\end{equation}
Putting $k=k_F|\sin(\theta/2)|$ and substituting the results of equations (\ref{barf1}) and (\ref{barf2}) into Eq.~(\ref{tildef1}) we find:
\begin{align}
\tilde F^{(1)}(\theta)&=\frac{4\pi\hbar^2 k_Fr_{*}}{m}|\sin\frac{\theta}{2}| +\frac{2\hbar^2}{m}(k_Fr_*)^2 \nonumber \\
&\times\pi\sin^2\frac{\theta}{2}\left[\ln|\xi r_{*}k_F\sin\frac{\theta}{2}|-\frac{25}{12}+3\ln 2\right]. \label{tildeF1}
\end{align}

The many-body correction (\ref{tildeE2}) we represent as $\tilde E^{(2)}=\tilde E_1^{(2)}+\tilde E_2^{(2)}$, where
\begin{align}
\!\!\!&\tilde E_1^{(2)}\!\!\!=\!\!-\frac{8(\pi\hbar r_*)^2}{mS^2}\!\!\!\!\!\sum_{\vec{k_{1}},\vec{k_{2}},\vec{k_{1}'}}\!\!\!\frac{|\vec{k'_1}\!-\!\vec{k_1}|^2}{\vec{k^{2}_{1}}\!+\!\vec{k^{2}_{2}}\!-\!\vec{k'^{2}_{1}}\!-\!\vec{k'^{2}_{2}}}n_\vec{k_{1}}n_\vec{k_{2}}n_\vec{k_{1}'}\!, \!\!\!\label{tildeE12} \\
\!\!\!&\tilde E_2^{(2)}\!\!\!=\!\frac{8(\pi\hbar r_*)^2}{mS^2}\!\!\!\!\!\sum_{\vec{k_{1}},\vec{k_{2}},\vec{k_{1}'}}\!\!\!\frac{|\vec{k_1}\!-\!\vec{k'_1}|\!\cdot\!|\vec{k_2}\!-\!\vec{k'_1}|}{\vec{k^{2}_{1}}\!+\!\vec{k^{2}_{2}}\!-\vec{k'^{2}_{1}}\!-\!\vec{k'^{2}_{2}}}n_\vec{k_{1}}n_\vec{k_{2}}n_\vec{k_{1}'}\!, \!\!\label{tildeE22}
\end{align}
and we used Eqs~(\ref{ffermion}) and (\ref{tildefBorn}) for the scattering amplitudes.
The contribution to the interaction function from $\tilde E_1^{(2)}$ is calculated in Appendix~\ref{App2} and it reads:
\begin{equation}   \label{tildeF12}
\!\!\!\tilde F_1^{(2)}\!(\theta)\!\!=\!\!\frac{2\hbar^2(k_Fr_*\!)^2}{m}\!\left[\!3\pi\!+\!2\pi\sin^2\frac{\theta}{2}\left(\!\frac{4}{3}\!-\!\ln\!|\tan \frac{\theta}{2}|\right)\right]\!.\!\!
\end{equation}
The contribution from $\tilde E_2^{(2)}$ is calculated in Appendix~\ref{F22}. It is given by 
\begin{widetext}
\begin{align}
\tilde{F}^{(2)}_2(\theta)=&\frac{2\hbar^2k^2_Fr^2_*}{m}\Big\{-\sin^2\frac{\theta}{2}\left(\pi\ln2+\frac{\pi}{2}-\pi\ln|\sin\frac{\theta}{2}|+4\ln|\cos\frac{\theta}{2}|-4\ln(1+|\sin\frac{\theta}{2}|)+\mathcal{G}(\theta)+
\frac{4\arcsin|\sin\frac{\theta}{2}|-2\pi}{|\cos\frac{\theta}{2}|} \right) \nonumber\\
&-\frac{1}{|\cos\frac{\theta}{2}|}\left(\pi-2\arcsin|\sin\frac{\theta}{2}|+|\sin\theta|\right)-4\left[\cos^2\frac{\theta}{2}\ln\frac{1+|\sin(\theta/2)|}{1-|\sin(\theta/2)|}+2|\sin\frac{\theta}{2}|\right]\Big\}, \label{tildeF22}
\end{align}
where
\begin{equation}  \label{G}
\mathcal{G}(\theta)=\int_{0}^{\pi} 2\sin^2\varphi \ln\left(\sin\varphi+\sqrt{\sin^2\frac{\theta}{2}+\cos^2\frac{\theta}{2}\sin^2\varphi}\right) d\varphi,
\end{equation}
so that
\begin{equation} \label{dGdtheta}
\frac{d\mathcal{G}(\theta)}{d\theta}=\frac{\pi}{2}\cot\frac{\theta}{2}-\frac{|\sin(\theta/2)|}{\sin(\theta/2)}\frac{1}{\cos(\theta/2)}+\frac{\arcsin|\cos(\theta/2)|}{|\cos(\theta/2)|}\left(\tan\frac{\theta}{2}-\cot\frac{\theta}{2}\right).
\end{equation}
\end{widetext}

We thus have $\tilde F(\theta)=\tilde{F}^{(1)}(\theta)+\tilde{F}^{(2)}_1(\theta)+\tilde{F}^{(2)}_2(\theta)$, where $\tilde F^{(1)}$, $\tilde F^{(2)}_1$, $\tilde F^{(2)}_2$ follow from  
Eqs.~(\ref{tildeF1}), (\ref{tildeF12}), and (\ref{tildeF22}). This allows us to proceed with the calculation of thermodynamic quantities.

\subsection{Compressibility, ground state energy, and effective mass}

We first calculate the compressibility at $T=0$. On the basis of Eq.~(\ref{u1}) we obtain: 
\begin{widetext}
\begin{align} \label{dmudN}
&\frac{\partial\mu}{\partial N}=\frac{2\pi\hbar^{2}}{mS}+\frac{1}{2\pi S}\int (1-\cos\theta) \left[\tilde{F}^{(1)}(\theta)+\tilde{F}_1^{(2)}(\theta)+\tilde{F}_2^{(2)}(\theta)\right] d\theta \nonumber\\
&=\frac{2\pi\hbar^2}{mS}+\frac{32\hbar^2}{3 mS}k_Fr_*+\frac{3\pi\hbar^2}{2mS}(k_Fr_*)^2\left(\ln[4\xi k_Fr_*]-\frac{3}{2}\right)+\frac{6\pi\hbar^2}{mS}(k_Fr_*)^2-\frac{\hbar^2}{\pi mS}(k_Fr_*)^2(30-8G+21\zeta(3)),
\end{align}
\end{widetext}
where $G=0.915966$ is the Catalan constant, and $\zeta(3)=1.20206$ is the Riemann zeta function. Calculating coefficients and recalling that $k_F=\sqrt{4\pi N/S}$ we represent the inverse compressibility following from Eq.~(\ref{kappa-1}) in a compact form:
\begin{equation}     \label{kappa*}
\!\!\kappa^{-1}\!=\!\frac{\hbar^2k_F^2}{2m}\frac{N}{S}\!\left(\!1\!+\!\frac{16}{3\pi}k_Fr_*\!+\!\frac{3}{4}(k_Fr_*)^2\ln(\zeta_1k_Fr_*)\right)\!,\!\!\!\!
\end{equation}
where we obtain the coefficient $\zeta_1=2.16\exp(-2A)$ by using Eq.~(\ref{xi}) for the coefficient $\xi$ which depends on the short-range behavior through the constant $A$ (see Eq.~(\ref{short})). For the chemical potential and ground state energy we obtain:
\begin{widetext}
\begin{eqnarray}
\mu&=&\frac{2\pi\hbar^2N}{mS}+\frac{64\hbar^2N}{9mS}k_Fr_*+\frac{3\pi\hbar^2N}{4mS}(k_Fr_*)^2\left(\ln[4\xi k_Fr_*]-\frac{7}{4}\right)+\frac{3\pi\hbar^2N}{mS}(k_Fr_*)^2-\frac{\hbar^2N}{2\pi mS}(k_Fr_*)^2(30-8G+21\zeta(3)) \nonumber \\
&=&\frac{\hbar^2k_F^2}{2m}\left(1+\frac{32}{9\pi}k_Fr_*+\frac{3}{8}(k_Fr_*)^2\ln(\zeta_2k_Fr_*)\right). \label{mug}
\end{eqnarray} 
\begin{eqnarray}
\frac{E}{N}&=&\frac{\pi\hbar^2N}{mS}+\frac{128\hbar^2N}{45mS}k_Fr_*+\frac{\pi\hbar^2N}{4mS}(k_Fr_*)^2\left(\ln[4\xi k_Fr_*]-\frac{23}{12}\right)+\frac{\pi\hbar^2N}{mS}-\frac{\hbar^2N}{6\pi mS}(30-8G+21\zeta(3)) \nonumber \\
&=&\frac{\hbar^2k_F^2}{4m}\left(1+\frac{128}{45\pi}k_Fr_*+\frac{1}{4}(k_Fr_*)^2\ln(\zeta_3k_Fr_*)\right), \label{Eg}
\end{eqnarray}
\end{widetext}
with numerical coefficients $\zeta_2=1.68\exp(-2A)$ and $\zeta_3=1.43\exp(-2A)$. Note that the first term in the second line of Eq.~(\ref{dmudN}) and the first terms in the first lines of Eqs.~(\ref{mug}) and (\ref{Eg}) represent the contributions of the kinetic energy, the second and third terms correspond to the contributions of the mean field part of the interaction energy, and the last two terms are the contributions of the many-body effects.    

The effective mass is calculated in a similar way by using Eq.~(\ref{eff}):
\begin{align}
\!\!\!\!\!\!&\frac{1}{m^*}\!=\!\frac{1}{m}\!-\!\frac{1}{(2\pi\hbar)^{2}}\int_{0}^{2\pi}\!(F^{(1)}(\theta)\!+\!F^{(2)}_1(\theta)\!+\!F^{(2)}_2(\theta))\cos\theta d\theta\!\!  \nonumber \\
\!\!\!\!\!\!&\!=\!\!\frac{1}{m}\!\!\left[\!1\!\!+\!\frac{4k_Fr_{\!*}}{3\pi}\!+\!\frac{(k_Fr_{\!*})^{\!2}}{4}\!\!\left(\!\ln{\![4k_Fr_{\!*}\xi]}\!\!-\!\frac{8}{3}\!+\!\frac{48G\!\!-\!\!20\!-\!\!14\zeta(3)}{\pi^2}\!\right)\!\right]\!\! \nonumber \\
\!\!\!\!\!\!&\!=\frac{1}{m}\left[1+\frac{4}{3\pi}k_Fr_*+\frac{1}{4}(k_Fr_*)^2\ln(\zeta_4k_Fr_*)\right],  \label{mg*}
\end{align}
where the numerical coefficient $\zeta_4=0.65\exp(-2A)$. Note that  if the potential $U(r)$ has the dipole-dipole form (\ref{VO}) up to very short distances, we have to put $A=0$ in the expressions for the coefficients $\zeta_1,\,\zeta_2,\,\zeta_3,\,\zeta_4$. Considering the quasi2D regime, this will be the case for $r_*$ greatly exceeding the length of the sample in the tightly confined direction, $l_0$. Then, as one can see from equations (\ref{kappa*}), 
(\ref{mug}), (\ref{Eg}), and (\ref{mg*}), the terms proportional to $(k_Fr_*)^2$ are always negative in the considered limit $k_Fr_*\ll 1$. These terms may become significant for $k_Fr_*>0.3$. 

\section{Zero sound}

In the collisionless regime of the Fermi liquid at very low temperatures, where the frequency of variations of the momentum distribution function greatly exceeds the relaxation rate of quasiparticles, one has zero sound waves. For these waves, variations $\delta n({\bf q},{\bf r},t)$ of the momentum distribution are related to deformations of the Fermi surface, which remains a sharp boundary between filled and empty quasiparticle states. At $T\rightarrow 0$ the equilibrium distribution $n_{\bf q}$ is the step function (\ref{nstep}), so that $\partial n_{\bf q}/\partial{\bf q}=-{\bf n}\delta(q-k_F)=-\hbar{\bf v}\delta(\epsilon_q-\epsilon_F)$, where ${\bf v}=v_F{\bf n}$, with ${\bf n}$ being a unit vector in the direction of ${\bf q}$. Then, searching for the variations $\delta n$ in the form:
$$\delta n({\bf q},{\bf r},t)=\delta(\epsilon_q-\epsilon_F)\nu({\bf n})\exp{i({\bf kr}-\omega t)}$$
and using Eq.~(\ref{1}), from the kinetic equation in the collisionless regime:
\begin{eqnarray}
\frac{\partial \delta n}{\partial t}+\vec{v}\cdot\frac{\partial \delta n}{\partial \vec{r}} -\frac{\partial n_{\bf q}}{\partial \vec{q}}\cdot\frac{\partial\delta\epsilon_q}{\hbar\partial \vec{r}}=0, \nonumber
\end{eqnarray}
one obtains an integral equation for the function $\nu({\bf n})$ representing displacements of the Fermi surface in the direction of ${\bf n}$ \cite{Landau}:
\begin{eqnarray}
(\omega-v_F\vec{n}\cdot\vec{k})\nu(\vec{n})=\frac{k_F}{(2\pi)^2\hbar}\vec{n}\cdot\vec{k}\int F(k_F\vec{n},k_F\vec{n'})\nu (\vec{n'})d{\bf n}'. \nonumber
\end{eqnarray}
Introducing the velocity of zero sound $u_0=\omega/k$ and dividing both sides of this equation by $v_Fk$ we have:
\begin{equation}  \label{inteq1}
(s-\cos\theta)\nu(\theta)=\frac{m^*\cos\theta}{(2\pi\hbar)^2}\int_0^{2\pi}\tilde F(\theta-\theta')\nu(\theta')d\theta',
\end{equation} 
where $s=u_0/v_F$, and $\theta,\,\theta'$ are the angles between ${\bf k}$ and ${\bf n},\,{\bf n}'$, so that $\theta-\theta'$ is the angle between ${\bf n}$ and ${\bf n}'$. The dependence of the interaction function of quasiparticles $\tilde F=\tilde F^{(1)}+\tilde F^{(2)}_1+\tilde F^{(2)}_2$ on $(\theta-\theta')$ follows from Eqs.~(\ref{tildeF1}), (\ref{tildeF12}), and (\ref{tildeF22}) in which one has to replace $\theta$ by $(\theta-\theta')$. 

The solution of equation (\ref{inteq1}) gives the function $\nu(\theta)$ and the velocity of zero sound $u_0$, and in principle one may obtain several types of solutions. It is important to emphasize that undamped zero sound requires the condition $s>1$, i.e. the sound velocity should exceed the Fermi velocity \cite{Landau}. We will discuss this issue below.  

For solving Eq.~(\ref{inteq1}) we represent the interaction function $\tilde F$ as a sum of the part proportional to $k_Fr_*$ and the part proportional to $(k_Fr_*)^2$. As follows from Eqs.~(\ref{tildeF1}), (\ref{tildeF12}), and (\ref{tildeF22}), we have:
\begin{equation}    \label{kandkk}
\!\!\!\!\tilde F(\theta\!-\!\theta')\!=\!\frac{4\pi\hbar^2}{m}k_Fr_*\!\left|\sin\frac{\theta\!-\!\theta'}{2}\right|+\frac{2\hbar^2}{m}(k_Fr_*\!)^2\Phi(\theta\!-\!\theta'),\!\!\!\! 
\end{equation} 
where the function $\Phi(\theta-\theta')$ is given by the sum of three terms. The first one is the term in the second line of Eq.~(\ref{tildeF1}), the second term is the expression in the square brackets in Eq.~(\ref{tildeF12}), the third term is the one in curly brackets in Eq.~(\ref{tildeF22}), and we should replace $\theta$ by $(\theta-\theta')$ in all these terms. It is important that the function $\Phi(\theta-\theta')$ does not have singularities and $\Phi(0)=\Phi(\pm 2\pi)=2\pi$. Using Eq.~(\ref{kandkk}) the integral equation (\ref{inteq1}) is reduced to the form:
\begin{align}     
(s-\cos\theta)\nu(\theta)&=\beta\cos\theta\int_0^{2\pi}\nu(\theta')\left|\sin\frac{\theta-\theta'}{2}\right|d\theta' \nonumber \\
&+\frac{\beta^2m}{2m^*}\cos\theta\int_0^{2\pi}\nu(\theta')\Phi(\theta-\theta')d\theta',\label{inteq2}
\end{align}
where $\beta=(m^*/\pi m)k_Fr_*\ll 1$. 

We now represent the function $\nu(\theta)$ as
\begin{equation}       \label{nucos}
\nu(\theta)=\sum_{p=0}^{\infty}C_p\cos p\theta.
\end{equation}
Then, integrating over $d\theta'$ in Eq.~(\ref{inteq2}), multiplying both sides of this equation by $\cos{j\theta}$ and integrating over $d\theta$, we obtain a system of linear equations for the coefficients $C_j$. We write this system for the coefficients $\eta_j=C_j(1-\beta/(j^2-1/4))$, so that $C_j=\eta_j(1+\beta U_j)$, where $U_j=(j^2-1/4-\beta)^{-1}$. The system reads:
\begin{align}   
&\!\!\!\!(s-1)(1+\beta U_0)\eta_0+[\eta_0-\frac{1}{2}\eta_1]+\beta U_0\eta_0=\frac{\beta^2}{2}{\bar \Phi}_0;\! \label{systemf0}\\
&\!\!\!\!(s\!-\!1)(1\!+\!\beta U_1)\eta_1\!+\![\eta_1-\eta_0-\frac{1}{2}\eta_2]+\beta U_1\eta_1\!\!=\!\frac{\beta^2}{2}{\bar \Phi}_1;\label{systemf1} \\
&\!\!\!\!(\!s\!\!-\!\!1\!)(\!1\!\!+\!\!\beta U_j\!)\eta_j\!\!+\!\![\eta_j\!\!-\!\!\frac{1}{2}\!(\!\eta_{j\!-\!1}\!\!+\!\eta_{j\!+\!1)\!}\!]\!\!+\!\!\beta U_j\eta_j\!\!=\!\!\frac{\beta^2}{2}\!{\bar \Phi}_j\!;\,j\!\!\geq\!\!2,\!\!\!\!\!\!\!\!\!\!\!\label{systemfj}
\end{align}
where
\begin{equation}    \label{Phij}
\!\!{\bar \Phi}_j\!=\!\frac{\tilde C_j}{\pi}\!\!\int_0^{2\pi}\!\!\!\!\!\!\!\!\cos\theta\,\cos j\theta d\theta\!\!\int_0^{2\pi}\!\!\sum_{p=0}^{\infty}C_p\cos p\theta'\Phi(\theta-\theta')d\theta'\!,\!\!\!\!\!\!\!
\end{equation}
with $\tilde C_j=1$ for $j\geq 1$ and $\tilde C_0=1/2$, and we put $m^*=m$ in the terms proportional to $\beta^2$.

In the weakly interacting regime the velocity of zero sound is close to the Fermi velocity and, hence, we have $(s-1)\ll 1$ (see, e.g. \cite{Landau}). Since $\beta\ll 1$, we first find coefficients $\eta_j$ omitting the terms proportional to $\beta$ and $\beta^2$ in Eqs.~(\ref{systemf0})-(\ref{systemfj}). For $j\gg 1$ equation (\ref{systemfj}) then becomes:
$$(s-1)\eta_j-\frac{1}{2}\frac{d^2\eta_j}{dj^2}=0,$$
and searching for $s>1$ we may write
\begin{equation}    \label{fjbig}
\eta_j\simeq\exp\{-\sqrt{2(s-1)}j\};\,\,\,j\gg 1.
\end{equation}
If $j\ll 1/\sqrt{s-1}$, then we may also omit the terms proportional to $(s-1)$ in the system of linear equations for $\eta_j$ (\ref{systemf0})-(\ref{systemfj}). The system then takes the form:
\begin{align}
&\eta_0-\frac{1}{2}\eta_1=0; \nonumber \\
&\eta_1-\eta_0-\frac{1}{2}\eta_2=0; \nonumber \\
&\eta_j-\frac{1}{2}[\eta_{j-1}+\eta_{j+1}]=0;\,\,\,\,j\geq 2. \nonumber
\end{align}
Without loss of generality we may put $\eta_0=1/2$. This immediately gives $\eta_j=1$ for $j\geq 1$, which is consistent with Eq.~(\ref{fjbig}) at $j\ll 1/\sqrt{s-1}$. We thus have the zero order solution:
\begin{equation}  \label{fzo}
\begin{cases}
\eta_0=1/2; \\
\eta_j=1;\,\,\,1\leq j\ll 1/\sqrt{s-1}.
\end{cases}
\end{equation}

In order to find the coefficients $\eta_j$ taking into account the terms linear in $\beta$, we consider $j$ such that $\beta U_j\sim\beta/j^2\gg (s-1)$, i.e. $j\ll\sqrt{\beta/(s-1)}$. Then we may omit the terms proportional to $(s-1)$ in equations (\ref{systemf0})-(\ref{systemfj}). Omitting also the terms proportional to $\beta^2$ this system of equations becomes:
\begin{align}
&\eta_0-\frac{1}{2}\eta_1+\beta U_0\eta_0=0;  \label{system0} \\
&\eta_1-\eta_0-\frac{1}{2}\eta_2+\beta U_1\eta_1=0; \label{system1} \\
&\eta_j-\frac{1}{2}[\eta_{j-1}+\eta_{j+1}]+\beta U_j\eta_j=0;\,\,j\geq 2. \label{systemj}
\end{align}
Putting again $\eta_0=1/2$ the solution of these equations reads:
\begin{align}
&\eta_1=1+\beta U_0; \nonumber \\
&\eta_j=1+\beta jU_0+2\beta\sum_{p=1}^{j-1}(j-p)U_p;\,\,\,j\geq 2. \nonumber
\end{align}
Confining ourselves to terms linear in $\beta$ we put $U_p=1/(p^2-1/4)$ and, hence, $U_0=-4$. Then, using the relation
$$\sum_{p=1}^{j-1}\frac{1}{p^2-1/4}=\frac{4(j-1)}{2j-1},$$
which is valid for $j\geq 2$, we obtain:
\begin{align} \label{fbeta}
\begin{cases}
&\eta_0=\frac{1}{2};   \\
\\
&\eta_1=1-4\beta;    \\
\\
&\eta_j=1-2\beta\left\{\frac{2j}{2j-1}+\sum_{p=1}^{j-1}\frac{p}{p^2-1/4}\right\};\,\,\,j\geq 2. 
\end{cases}
\end{align}
For $j\gtrsim\sqrt{\beta/(s-1)}$ we should include the terms proportional to $(s-1)$ in Eq.~(\ref{systemj}). This leads to the solution in the form of the decaying Bessel function: 
$\eta_j\simeq\sqrt{2(s-1)/\pi}K_{\sqrt{1/4+\beta}}(\sqrt{s-1}j)$, which for small $\beta$ is practically equivalent to Eq.~(\ref{fjbig}).

We now make a summation of equations (\ref{systemf0})-(\ref{systemfj}) from $j=0$ to $j=j_*\ll1/\sqrt{s-1}$. The summation of the second terms of these equations gives $\sqrt{(s-1)/2}$, whereas the contribution of the terms proportional to $(s-1)$ is much smaller and will be omitted. The sums $\sum_{j=0}^{j_*}U_j\eta_j$ and $\sum_{j=0}^{j_*}{\bar \Phi}_j$ converge at $j\ll 1/\sqrt{s-1}$, and the upper limit of summation in these terms can be formally replaced by infinity. We thus obtain a relation:
\begin{equation}   \label{sum1}
\sqrt{\frac{s-1}{2}}+\sum_{j=0}^{\infty}\beta\eta_jU_j-\frac{\beta^2}{2}\sum_{j=0}^{\infty}{\bar \Phi}_j=0.
\end{equation}
Confining ourselves to contributions up to $\beta^2$, in the second term on the left hand side of Eq.~(\ref{sum1}) we use coefficients $\eta_j$ given by Eqs.~(\ref{fbeta}), and write $U_j=1/(j^2-1/4)+\beta/(j^2-1/4)^2$. In the expressions for ${\bar \Phi}_j$ we use $C_0=1/2$ and $C_p=1$ for $p\geq 1$. We then have:
\begin{equation}
\!\!\sum_{j=0}^{\infty}\beta\eta_jU_j\!=\!-2\beta\!+\!\sum_{j=1}^{\infty}\frac{\beta}{\!j^2\!-\!1/4}\!+\!\beta^2\{\!8\!+\!S_1\!-\!2S_2\!-2S_3\!\}. \!\!\!\!\label{sum2}
\end{equation}
The contribution linear in $\beta$ vanishes because $\sum_{j=1}^{\infty}1/(j^2-1/4)=2$. The quantities $S_1,\,S_2$, and $S_3$ are given by
\begin{align}
&\!\!\!S_1=\sum_{j=1}^{\infty}\frac{1}{(j^2-1/4)^2}=\pi^2-8;  \nonumber \\
&\!\!\!S_2\!=\!\sum_{j=2}^{\infty}\frac{1}{j^2\!-\!1/4}\!\sum_{p=1}^{j-1}\frac{p}{p^2\!-\!1/4}\!=\!\sum_{j=1}\frac{j}{(j^2\!-\!1/4)(j\!+\!1/2)}; \nonumber \\
&\!\!\!S_3=\sum_{j=1}^{\infty}\frac{j}{(j-1/2)(j^2-1/4)}, \nonumber
\end{align}
so that
$$S_2+S_3=\sum_{j=1}^{\infty}\frac{2j^2}{(j^2-1/4)^2}=\frac{\pi^2}{2}.$$
We thus see that the contribution quadratic in $\beta$ also vanishes because the term in the curly brackets in Eq.~(\ref{sum2}) is exactly equal to zero. Hence, we have $\sum_{j=0}^{\infty}\beta\eta_jU_j=0$ up to terms proportional to $\beta^2$.

The sum in the third term on the left hand side of Eq.~(\ref{sum1}), after putting $C_0=1/2$ and $C_p=1$ for $p\geq 1$ in the relations for ${\bar \Phi}_j$, reduces to
\begin{align}    \label{sum3}
\sum_{j=0}^{\infty}{\bar \Phi}_j&=\frac{1}{4\pi}\!\sum_{j=-\infty}^{\infty}\!\int_0^{2\pi}\!\!\!\!\cos\theta\,\cos{j\theta}d\theta  \nonumber \\
&\times\sum_{p=-\infty}^{\infty}\!\int_0^{2\pi}\!\!\!\!\cos{p\theta'}\Phi(\theta\!-\!\theta')d\theta'.\!\!\!\!
\end{align}   
For $\theta$ in the interval $0\leq\theta\leq 2\pi$ we have a relation:
$$\sum_{j=-\infty}^{\infty}\cos{j\theta}=\pi[\delta(\theta)+\delta(\theta-2\pi)],$$
which transforms Eq.~(\ref{sum3}) to
\begin{equation}   \label{sum4}
\sum_{j=0}^{\infty}{\bar \Phi}_j=\frac{\pi}{4}[2\Phi(0)+\Phi(2\pi)+\Phi(-2\pi)]=2\pi^2,
\end{equation}
and equation (\ref{sum1}) becomes:
\begin{eqnarray}
\sqrt{\frac{s-1}{2}}-\beta^2\pi^2=0. \nonumber
\end{eqnarray}
This gives $s=1+2(\beta\pi)^4$, and recalling that $\beta=k_Fr_*/\pi$ (we put $m^*=m$) we obtain for the velocity of zero sound:
\begin{equation}   \label{u0}
u_0=v_F[1+2(k_Fr_*)^4].
\end{equation} 

Note that in contrast to the 3D two-species Fermi gas with a weak repulsive contact interaction (scattering length $a$), where the correction $(u_0-v_F)$ exponentially depends on $k_Fa$, for our 2D dipolar gas we obtained a power law dependence. This is a consequence of dimensionality of the system. 

It is important that confining ourselves to only the leading part of the interaction function $\tilde F$, which is proportional to $k_Fr_*$ and is given by the first term of Eq.~(\ref{tildeF1}), we do not obtain undamped zero sound ($s>1$) \cite{comment}. This corresponds to omitting the terms $\beta^2{\bar \Phi}_j/2$ in equations (\ref{systemf0})-(\ref{systemfj}) and is consistent with numerical calculations \cite{Baranov}. Only the many-body corrections to the interaction function of quasiparticles, given by equations (\ref{tildeF12}) and (\ref{tildeF22}), provide non-zero positive values of $\Phi(0)$ and $\Phi(\pm 2\pi)$, thus leading to a positive value of $(u_0-v_F)$. One then sees that many-body effects are crucial for the propagation of zero sound.

In principle, we could obtain the result of Eq.~(\ref{u0}) in a simpler way, similar to that used for the two-species Fermi gas with a weak repulsive interaction (see, e.g. \cite{Landau}). Representing the function $\nu(\theta)$ as $\nu(\theta)=\cos{\theta}\tilde\nu(\theta)/(s-\cos{\theta})$ we transform Eq.~(\ref{inteq1}) to the form:
\begin{equation}   \label{inteq3}
\tilde\nu(\theta)=\frac{m^*}{(2\pi\hbar)^2}\int_0^{2\pi}\frac{\tilde F(\theta-\theta')\tilde\nu(\theta')\cos{\theta'}}{s-\cos{\theta'}}d\theta'.
\end{equation}    
Since $s$ is close to unity, it looks reasonable to assume that the main contribution to the integral in Eq.~(\ref{inteq3}) comes from $\theta'$ close to zero and to $2\pi$. Using the fact that $\tilde F(\theta)=\tilde F(2\pi -\theta)$ we then obtain:
\begin{equation}   \label{tildenu}
\tilde\nu(\theta)=\frac{m^*\tilde F(\theta)\tilde\nu(0)}{4\pi\hbar^2}\sqrt{\frac{2}{s-1}}.
\end{equation}
We now take the limit $\theta\rightarrow 0$ and substitute $\tilde F(0)=(4\pi\hbar^2/m)(k_Fr_*)^2$ as follows froms Eqs.~(\ref{tildeF1}), (\ref{tildeF12}), and (\ref{tildeF22}). Putting $m^*=m$ we then obtain $s=1+2(k_Fr_*)^4$ and arrive at Eq.~(\ref{u0}). 

Note, however, that for very small $\theta$ or $\theta$ very close to $2\pi$ the dependence $\tilde F(\theta)$ is very steep. For $\theta\rightarrow 0$ the leading part of the interaction function, which is linear in $k_Fr_*$, vanishes, and only the quadratic part contributes to $\tilde F(0)$. Therefore, strictly speaking the employed procedure of calculating the integral in Eq.~(\ref{inteq3}) is questionable for very small $\theta$. This prompted us to make the analysis based on representing $\nu(\theta)$ in the form (\ref{nucos}) and on solving the system of linear equations (\ref{systemf0})-(\ref{systemfj}). 

Equation (\ref{inteq3}) is useful for understanding why undamped zero sound requires the condition $s>1$ so that $u_0>v_F$. For $s<1$ there is a pole in the integrand of Eq.~(\ref{inteq3}), which introduces an imaginary part of the integral. As a result, the zero sound frequency $\omega$ will also have an imaginary part at real momenta $k$, which means the presence of damping (see, e.g. \cite{Landau}).

We could also consider an odd function $\nu(\theta)$, namely such that $\nu(2\pi-\theta)=-\nu(\theta)$ and $\nu(0)=\nu(2\pi)=0$. In this case, however, we do not obtain an undamped zero sound.

\section{Concluding remarks}

We have shown that (single-component) fermionic polar molecules in two dimensions constitute a novel Fermi liquid, where many-body effects play an important role. For dipoles oriented perpendicularly to the plane of translational motion, the many-body effects provide significant corrections to thermodynamic functions. Revealing these effects is one of the interesting goals of up-coming experimental studies. The investigation of the full thermodynamics of 2D polar molecules, including many-body effects, can rely on the in-situ imaging technique as it has been done for two-component atomic Fermi gases \cite{Salomon1,Salomon2}. This method can also be extended to 2D systems for studying thermodynamic quantities \cite{Zwierlein, Dalibard2}.  Direct imaging of a 3D pancake-shaped dipolar molecular system has been recently demonstrated at JILA \cite{Ye2}. For 2D polar molecules discussed in our paper, according to equations (\ref{kappa*})-(\ref{Eg}), the contribution of many-body corrections proportional to $(k_Fr_*)^2$ can be on the level of $10\%$ or $20\%$ for $k_Fr_*$ close to $0.5$. Thus, finding many-body effects in their thermodynamic properties looks feasible.   

It is even more important that the many-body effects are responsible for the propagation of zero sound waves in the collisionless regime of the 2D Fermi liquid of polar molecules with dipoles perpendicular to the plane of translational motion. This is shown in Section IV of our paper, whereas mean-field calculations do not find undamped zero sound \cite{Baranov}. Both collisionless and hydrodynamic regimes are achievable in on-going experiments. This is seen from the dimensional estimate of the relaxation rate of quasiparticles. At temperatures $T\ll \epsilon_F$ the relaxation of a non-equilibrium distribution of quasiparticles occurs due to binary collisions of quasiparticles with energies in a narrow interval near the Fermi surface. The width of this interval is $\sim T$ and, hence, the relaxation rate contains a small factor $(T/\epsilon_F)^2$ (see, e.g. \cite{Landau}). Then, using the Fermi Golden rule we may write the inverse relaxation time as $\tau^{-1}\sim (g_{eff}^2/\hbar)(m/\hbar^2)n(T/\epsilon)^2$, where $n$ is the 2D particle density, the quantity $\sim m/\hbar^2$ represents the density of states on the Fermi surface, and the quantity $g_{eff}$ is the effective interaction strength. Confining ourselves to the leading part of this quantity, from Eqs.~(\ref{1str2}) and (\ref{barf1}) we have $g_{eff}\sim \hbar^2k_Fr_*/m$. We thus obtain:
\begin{equation}   \label{taurel}
\frac{1}{\tau}\sim\frac{\hbar n}{m}(k_Fr_*)^2\left(\frac{T}{\epsilon_F}\right)^2.
\end{equation}
Note that as $\epsilon_F\approx\hbar^2k_F^2/2m\approx 2\pi\hbar^2n/m$, for considered temperatures $T\ll \epsilon_F$ the relaxation time $\tau$ is density independent. 
Excitations with frequencies $\omega\ll 1/\tau$ are in the hydrodynamic regime, where on the length scale smaller than the excitation wavelength and on the time scale smaller than $1/\omega$ the system reaches a local equilibrium. On the other hand, excitations with frequencies $\omega\gg 1/\tau$ are in the collisionless regime. Assuming $T\sim 10$nK, for KRb molecules characterized by the dipole moment $d\simeq 0.25$ D in the electric field of $5$kV/cm as obtaind in the JILA experiments, we find $\tau$ on the level of tens of milliseconds. The required condition $T\ll\epsilon_F$ is satisfied for $\epsilon_F\gtrsim 70$ nK, which corresponds to $n\gtrsim 2\cdot 10^8$ cm$^{-2}$. In such conditions excitations with frequencies of the order of a few Hertz or lower will be in the hydrodynamic regime, and excitations with larger frequencies in the collisionless regime.    

The velocity of zero sound is practically equal to the Fermi velocity $v_F=\hbar k_F/m^*$. This is clearly seen from Eq.~(\ref{u0}) omitting a small correction proportional to $(k_Fr_*)^4$. Then, using Eq.~(\ref{mg*}) for the effective mass and retaining only corrections up to the first order in $k_Fr_*$, we have:
\begin{equation}   \label{u0Simple}
u_0\simeq\frac{\hbar k_F}{m}\left(1+\frac{4}{3\pi}k_Fr_*\right).
\end{equation}
In the hydrodynamic regime the sound velocity is:
\begin{equation}    \label{usimple}
u=\sqrt{\frac{N}{m}\frac{\partial\mu}{\partial N}}\simeq\frac{\hbar k_F}{m}\left(1+\frac{8}{3\pi}k_Fr_*\right),
\end{equation}
where we used Eq.~(\ref{dmudN}) for $\partial\mu/\partial N$ and retained corrections up to the first order in $k_Fr_*$. The hydrodynamic velocity $u$ is slightly larger than the velocity of zero sound $u_0$, and the difference is proportional to the interaction strength. This is in sharp contrast with the 3D two-component Fermi gas, where $u_0\approx v_F>u\approx v_F/\sqrt{3}$. 

We thus see that it is not easy to distinguish between the hydrodynamic and collisionless regimes from the measurement of the sound velocity. A promising way to do so can be the observation of damping of driven excitations, which in the hydrodynamic regime is expected to be slower. Another way is to achieve the values of $k_Fr_*$ approaching unity and still discriminate between $u_0$ and $u$ in the measurement of the sound velocity. For example, in the case of dipoles perpendicular to the plane of their translational motion the two velocities are different from each other by about $20\%$ at $k_Fr_*\simeq 0.5$. These values of $k_Fr_*$ are possible if the 2D gas of dipoles still satisfies the Pomeranchuk criteria of stability. These criteria require that the energy of the ground state corresponding to the occupation of all quasiparticle states inside the Fermi sphere, remains the minimum energy under an arbitrarily small deformation of the Fermi sphere. The generalization of the Pomeranchuk stability criteria to the case of the 2D single-component Fermi liquid with dipoles perpendicular to the plane of their translational motion reads:
\begin{equation}   \label{Pom}
1+\frac{m^*}{(2\pi\hbar)^2}\int_0^{2\pi}\tilde F(\theta)\cos{j\theta}\,d\theta>0,
\end{equation}
and this inequality should be satisfied for any integer $j$. As has been found in Ref.~\cite{Baranov}, the Pomeranchuk stability criteria (\ref{Pom}) are satisfied for $k_Fr_*$ approaching unity from below if the interaction function of quasiparticles contains only the first term of Eq.~(\ref{tildeF1}), which is the leading mean field term. We have checked that the situation with the Pomeranchuk stability does not change when we include the full expression for the interaction function, $\tilde F(\theta)=\tilde F^{(1)}(\theta)+\tilde F^{(2)}_1(\theta)+\tilde F^{(2)}_2(\theta)$, following from Eqs.~(\ref{tildeF1}), (\ref{tildeF12}), and (\ref{tildeF22}). Thus, achieving $k_Fr_*$ approaching unity looks feasible. For KRb molecules with the (oriented) dipole moment of $0.25$ D the value $k_Fr_*\approx 0.5$ requires densities $n\approx 2\cdot 10^8$ cm$^{-2}$.

Finally, we would like to emphasize once more that our results are applicable equally well for the quasi2D regime, where the dipole-dipole length $r_*$ is of the order of or smaller than the confinement length $l_0=(\hbar/m\omega_0)^{1/2}$, with $\omega_0$ being the frequency of the tight confinement. The behavior at distances $r\lesssim l_0$ is contained in the coefficient $A$ defined in Eq.~(\ref{short}). Therefore, the results for the velocity of zero sound which is independent of $A$, are universal in the sense that they remain unchanged when going from $r_*\gg l_0$ to $r_*\lesssim l_0$. The only requirement is the inequality $k_Fl_0\ll 1$. It is, however, instructive to examine the ratio $r_*/l_0$ that can be obtained in experiments with ultracold polar molecules. Already in the JILA experiments using the tight confinement of KRb molecules with frequency $\omega_0\approx 30$ kHz and achieving the average dipole moment $d\simeq 0.25$ D in electric fields of $5$ kV/cm, we have $r_*\simeq 100$ nm and $l_0\simeq 50$ nm so that $r_*/l_0\simeq 2$. A decrease of the confinement frequency to $5$ kHz and a simultaneous decrease of the dipole moment by a factor of 2 leads to $r_*/l_0\sim 0.2$. On the other hand, for $d$ close to $0.5$ (which is feasible to obtain for other molecules) one can make the ratio $r_*/l_0$ close to $10$ at the same confinement length.    
     
\section*{Acknowledgements}

We are grateful to M.A. Baranov and S.I. Matveenko for fruitful discussions. We acknowledge support from EPSRC Grant No. EP/F032773/1, from the IFRAF Institute, and from the Dutch Foundation FOM. This research has been supported in part by the National Science Foundation under Grant No. NSF PHYS05-51164. LPTMS is a mixed research unit No. 8626 of CNRS and Universit\'e Paris Sud.

\appendix
\section{Direct calculation of the first order contribution to the interaction energy} 
\label{App1}
For directly calculating the first order (mean field) contribution to the interaction energy $\tilde E^{(1)}$ (\ref{tildeE1}), we represent it as $\tilde E^{(1)}=\tilde E^{(1)}_1+\tilde E^{(1)}_2$ where
\begin{align}
&\tilde E^{(1)}_1=\int {\bar f}^{(1)}\left(\frac{|{\bf k}_1-{\bf k}_2|}{2}\right)n_{{\bf k}_1}n_{{\bf k}_2}\frac{d^2k_1d^2k_2}{(2\pi)^4},\label{E11} \\
&\tilde E^{(1)}_2=\int {\bar f}^{(2)}\left(\frac{|{\bf k}_1-{\bf k}_2|}{2}\right)n_{{\bf k}_1}n_{{\bf k}_2}\frac{d^2k_1d^2k_2}{(2\pi)^4},\label{E12} 
\end{align}
and the amplitudes  ${\bar f}^{(1)}$ and  ${\bar f}^{(2)}$ are given by Eqs.~(\ref{barf1}) and (\ref{barf2}), respectively. In the calculation of the integrals for $\tilde E^{(1)}_1$ and $\tilde E^{(1)}_2$ we turn to the variables ${\bf x}=({\bf k}_1-{\bf k}_2)/2k_F$ and ${\bf y}=({\bf k}_1+{\bf k}_2)/2k_F$, so that $d^2k_1d^2k_2=8\pi k_F^4d^2xd^2yd\varphi$, where $\varphi$ is the angle between the vectors ${\bf x}$ and ${\bf y}$, and the integration over $d\varphi$ should be performed from $0$ to $2\pi$. The distribution functions $n_{{\bf k}_1}$ and $n_{{\bf k}_2}$ are the step functions (\ref{nstep}). The integration over $dk_1$ and $dk_2$ from $0$ to $k_F$ corresponds to the integration over $dy$ from $0$ to $y_0(x,\varphi)=-x|\cos\varphi|+\sqrt{1-x^2\sin2\varphi}$ and over $dx$ from $0$ to $1$. Using Eq.~(\ref{barf1}) we reduce Eq.~(\ref{E11}) to 
\begin{equation} \label{intermE11}
\tilde E^{(1)}_1=\frac{S\hbar^2k_F^4}{\pi^2m}k_Fr_*I_1,
\end{equation}
where
\begin{align}
&I_1=\int_0^{2\pi}d\varphi\int_0^1x^2dx\int_0^{y_0(x,\varphi)}ydy=\frac{1}{2}\int_0^{2\pi}d\varphi\int_0^1x^2 dx\nonumber\\
&\times[1-2|\cos\varphi|\sqrt{1-x^2\sin^2\varphi}+x^2(\cos^2\varphi-\sin^2\varphi)].\nonumber
\end{align}
The last term of the second line vanishes, and the integration of the first two terms over $d\varphi$ and $dx$ gives:
\begin{equation*}
I_1=\int_0^1x^2\left(\pi-2x\sqrt{1-x^2}-2\arcsin{x}\right)=\frac{8}{45}.
\end{equation*}
Then Eq.~(\ref{intermE11}) yields:
\begin{equation}    \label{E11final}
\tilde E^{(1)}_1=\frac{8S}{45\pi^2}\frac{\hbar^2k_F^4}{m}k_Fr_*=\frac{N^2}{S}\frac{128}{45}\frac{\hbar^2k_F^2}{m}k_Fr_*,
\end{equation}
which exactly coincides with the second term of the first line of Eq.~(\ref{Eg}).

Using Eq.~(\ref{barf2}) the contribution $\tilde E^{(1)}_2$ takes the form:
\begin{equation} \label{E12interm}
\!\!\!\!\tilde E^{(\!1\!)}_2\!\!\!=\!\frac{S\hbar^2\!k_F^4}{2\pi^2m}(k_Fr_*\!)^2\!\left\{\!\left[\ln(\xi k_Fr_*\!)\!-\!\frac{25}{12}\!+\!3\ln{2}\right]\!I_2\!+\!I_3\!\right\}\!\!,\!\!\!\!\!\!\!\!
\end{equation}
where the integrals $I_2$ and $I_3$ are given by
\begin{align}
I_2&=\int_0^{2\pi}d\varphi\int_0^1x^3dx\int_0^{y_0(x,\varphi)}ydy =\frac{1}{2}\int_0^{2\pi}d\varphi\int_0^1x^3 dx\nonumber\\
&\times\big[1-2|\cos\varphi|x\sqrt{1-x^2\sin^2\varphi}+x^2(\cos^2\varphi-\sin^2\varphi)\big] \nonumber \\
&=\frac{1}{2}\int_0^1x^3[2\pi-4x\sqrt{1-x^2}-4\arcsin{x}]dx=\frac{\pi}{32},\nonumber
\end{align}
and
\begin{align}
&I_3=\int_0^{2\pi}d\varphi\int_0^1x^3\ln{x}dx\int_0^{y_0(x,\varphi)}ydy=\frac{1}{2}\int_0^{2\pi}d\varphi\int_0^1 dx\nonumber\\
&\times x^3\ln{x}\big[1-2|\cos\varphi|x\sqrt{1-x^2\sin^2\varphi}+x^2(\cos^2\varphi-\sin^2\varphi)\big]\nonumber \\
&=\frac{1}{2}\int_0^1x^3\ln{x}[2\pi-4x\sqrt{1-x^2}-4\arcsin{x}]dx\nonumber\\
&=\frac{\pi}{32}\left(\frac{1}{6}-\ln{2}\right).\nonumber
\end{align}

Substituting the calculated $I_2$ and $I_3$ into Eq.~(\ref{E12interm}) we obtain:
\begin{align}  
&\tilde E^{(1)}_2=\frac{S\hbar^2k_F^4}{64\pi m}(k_Fr_*)^2\left[\ln(4\xi k_Fr_*)-\frac{23}{12}\right] \nonumber \\
&=\frac{N^2}{S}\frac{\pi\hbar^2}{4m}(k_Fr_*)^2\left[\ln(4\xi k_Fr_*)-\frac{23}{12}\right]. \label{E12final}
\end{align}
This exactly reproduces the third term of the first line of Eq.~(\ref{Eg}).

\section{Calculation of the interaction function $\tilde F_{1}^{(2)}$} 
\label{App2}
The interaction function $\tilde F_{1}^{(2)}$ is the second variational derivative of the many-body contribution to the interaction energy, $\tilde E^{(2)}_1$ (\ref{tildeE12}), with respect to the momentum distribution function. It can be expressed as 
\begin{equation}  \label{F12initial}
\tilde F_{1}^{(2)}(\vec{k},\vec{k'})=-\frac{2\hbar^2}{m}(k_Fr_*)^2 (\tilde I_1+\tilde I_2+\tilde I_3), 
\end{equation}
where
\begin{align}
\!\!&\tilde I_1=2\!\int_{|\vec{k_1}|<k_F}\!\frac{d^2k_1}{k_F^2}\frac{|\vec{k}-\vec{k_1}|^2}{\vec{k^2}\!+\!\vec{k'^2}\!-\!\vec{k^2_1}\!-\!\vec{k^2_2}}\delta_{\vec{k}\!+\!\vec{k'}\!-\!\vec{k_1}\!-\!\vec{k_2}},
\label{tildeI1} \\
\!\!&\tilde I_2=2\!\int_{|\vec{k_1}|<k_F}\!\frac{d^2k_1}{k_F^2}\frac{|\vec{k}-\vec{k'}|^2}{\vec{k^2}\!+\!\vec{k^2_1}\!-\!\vec{k'^2}\!-\!\vec{k^2_2}}\delta_{\vec{k}\!+\!\vec{k_1}\!-\!\vec{k'}\!-\!\vec{k_2}},
\label{tildeI2} \\
\!\!&\tilde I_3=2\!\int_{|\vec{k_1}|<k_F}\!\frac{d^2k_1}{k_F^2}\frac{|\vec{k_1}-\vec{k'}|^2}{\!\vec{k^2_1}\!+\!\vec{k^2}\!-\!\vec{k'^2}\!-\!\vec{k^2_2}}\delta_{\vec{k_1}\!+\!\vec{k}\!-\!\vec{k'}\!-\!\vec{k_2}},
\label{tildeI3}
\end{align}
and the presence of the Kronecker symbols $\delta_{\bf q}$ reflects the momentum conservation law. On the Fermi surface we put $|{\bf k}|=|{\bf k}'|=k_F$ and denote the angle between ${\bf k}$ and ${\bf k}'$ as $\theta$.  Due to the symmetry property: $F(\vec{k},\vec{k'})=F(\vec{k'},\vec{k})$ we have $F(\theta)=F(2\pi-\theta)$ and may consider $\theta$ in the interval from $0$ to $\pi$. 

\begin{figure}[ttp]
\includegraphics[width=0.5\columnwidth]{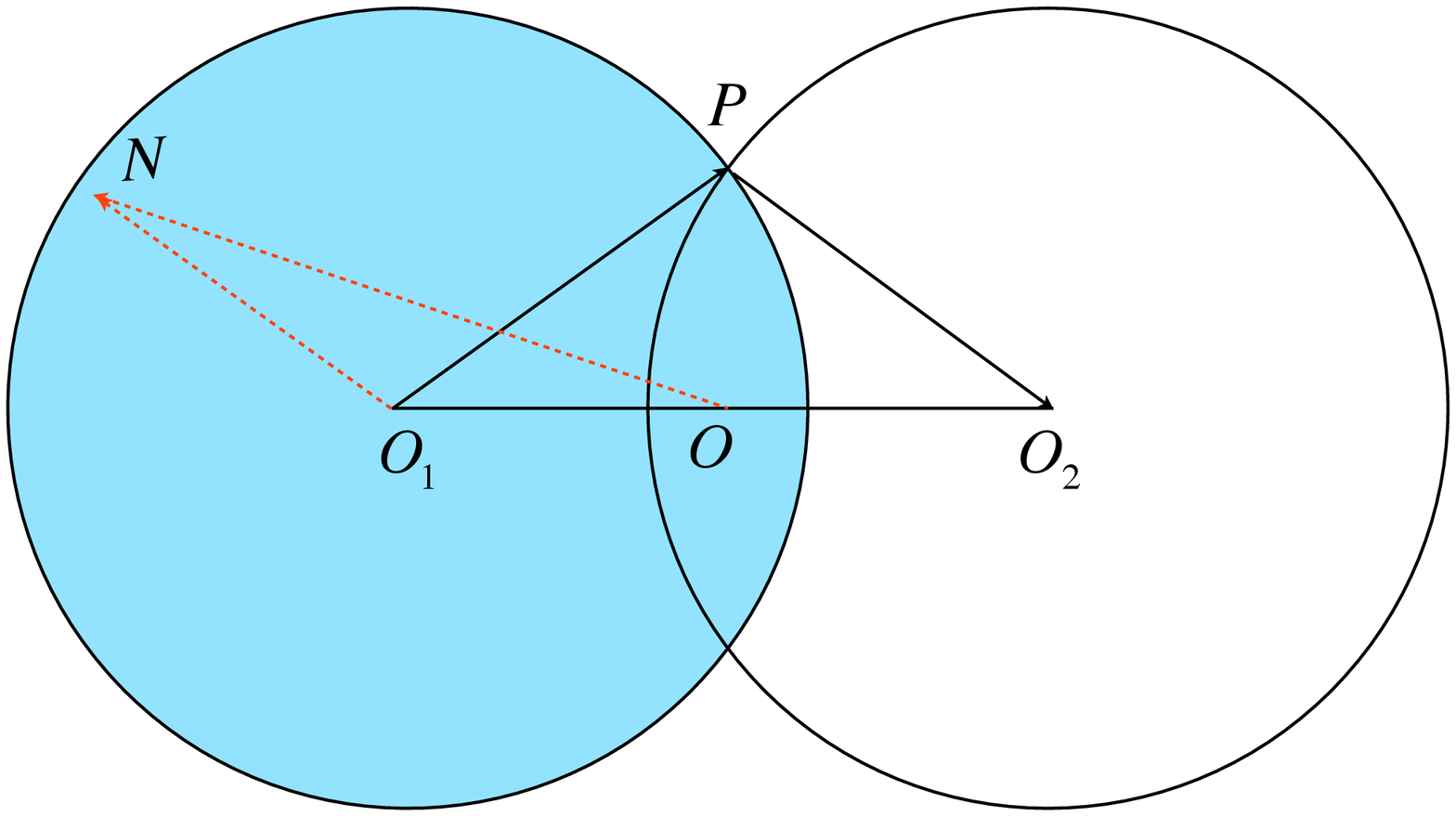}\hfill
\includegraphics[width=0.5\columnwidth]{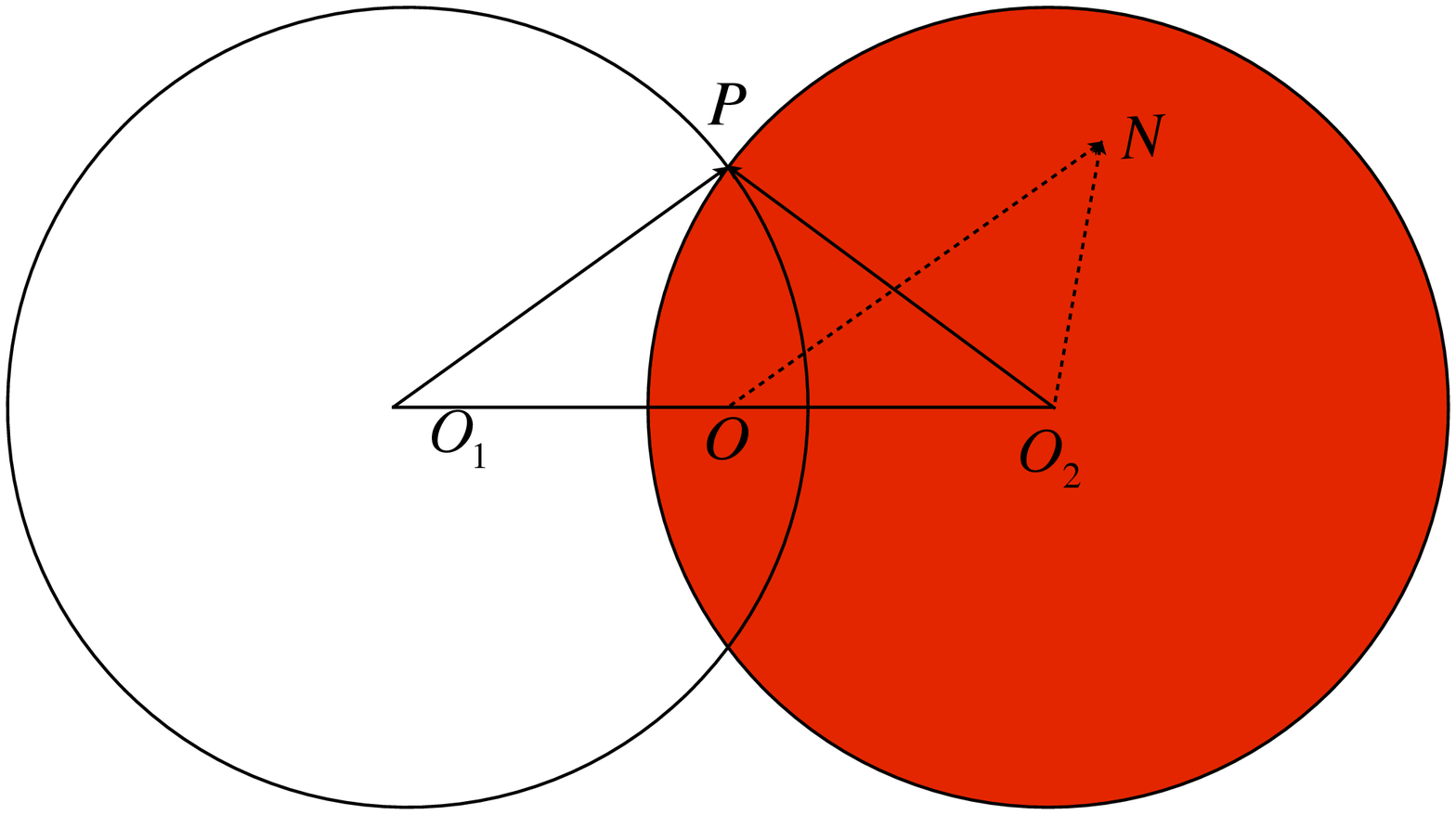}
\caption{(color online).
Left: The integration area for $\tilde I_1$ (in blue). The distance between the points $O_1$ and $P$ is ${\bf R}_{O_1P}=\vec{k}/2k_F$, ${\bf R}_{PO_2}=\vec{k'}/2k_F,$ and ${\bf R}_{O_1N}=\vec{k_1}/2k_F$. Right: The integration area for $\tilde I_2$ and $\tilde I_3$ (in red). The distance between the points $O_2$ and $N$ is ${\bf R}_{O_2N}=\vec{k_1}/2k_F$, ${\bf R}_{O_1P}={\bf k}/2k_F$, and ${\bf R}_{O_2P}={\bf k}'/2k_F$.
\label{fig:f1}}
\end{figure}

In order to calculate the integral $\tilde I_1$, we use the quantities $\vec{s}=(\vec{k}+\vec{k'})/2k_F$ and $\vec{m}=(\vec{k}-\vec{k'})/2k_F$ and turn to the variable $\vec{x}=(\vec{k_1}-\vec{k_2})/2k_F=(2\vec{k_1}-\vec{s})/2k_F$. For given vectors $\vec{k}$ and $\vec{k'}$, the vectors $\vec{s}$ and $\vec{m}$ are fixed and $|\vec{s}|=\cos(\theta/2)$, $|\vec{m}|=\sin(\theta/2)$.  The integral can then be rewritten as:
\begin{equation*}
\tilde I_1=\int \frac{m^2+x^2}{m^2-x^2} d^2x.
\end{equation*}
The integration region is shown in Fig.\ref{fig:f1}, where the distance between the points $O_1$ and $O_2$ is ${\bf R}_{O_1O_2}=\vec{s}$. The distance between the points $O_1$ and $N$ is ${\bf R}_{O_1N}=\vec{k_1}/2k_F$, and ${\bf R}_{NO_2}=\vec{k_2}/2k_F$, so that ${\bf R}_{ON}=\vec{x}/2$. The quantity $|{\bf x}|$ changes from $0$ to $l_1(\varphi)$ where 
$$l_1^2(\varphi)+\cos^2\frac{\theta}{2}-2l_1(\varphi)\cos\frac{\theta}{2}\cos\varphi=1,$$ 
and $l_1(\varphi)\cdot l_1(\varphi+\pi)=\sin^2(\theta/2)$, with $\varphi$ being an angle between ${\bf m}$ and ${\bf x}$. In the polar coordinates the integral $\tilde I_1$ takes the form:
\begin{equation*}
\tilde I_1=\int_0^{2\pi}d\varphi\int_0^{l_1(\varphi)}\left(-1+2\sin^2\frac{\theta}{2}\frac{1}{\sin^2(\theta/2)-x^2}\right)xdx,
\end{equation*}
and after a straightforward integration we obtain:
\begin{equation} \label{tildeI1final}
\tilde I_1=\pi\left(2\sin^2\frac{\theta}{2}\ln|\tan\frac{\theta}{2}|-1\right).
\end{equation}

In the integral $\tilde I_2$, using the variable $\vec{y}=(\vec{k_1}+\vec{k_2})/2k_F$ we observe that it changes from $0$ to $l_2(\tilde\varphi)$ where 
$$l_2^2(\tilde\varphi)+\sin^2\frac{\theta}{2}-2l_2(\tilde\varphi)\sin\frac{\theta}{2}\cos(\tilde\varphi)=1$$ 
and $l_2(\tilde\varphi)-l_2(\tilde\varphi+\pi)=2\sin\frac{\theta}{2}\cos\tilde\varphi$, with $\tilde\varphi$ being an angle between ${\bf y}$ and ${\bf m}$. We then have:
\begin{align}
\tilde I_2&=-2\int \frac{m^2}{\vec{m}\cdot\vec{y}}d^2y=-2\sin\frac{\theta}{2}\int_{0}^{2\pi}d\tilde\varphi\int_0^{l_2(\tilde\varphi)} \frac{dy}{\cos\tilde\varphi}\nonumber \\
&=-4\pi\sin^2\frac{\theta}{2}. \label{tildeI2final}
\end{align} 

For the integral $\tilde I_3$ we have:
\begin{align}
&\tilde I_3=-\frac{1}{2}\int \frac{s^2+y^2-2\vec{s}\cdot\vec{y}}{\vec{m}\cdot\vec{y}} \nonumber\\
&=-\frac{1}{2\sin\frac{\theta}{2}}\int_{0}^{2\pi}\!\!\frac{d\tilde\varphi}{\cos\tilde\varphi}\!\int_{0}^{l_2(\tilde\varphi)}\!\!\!dy\left[y^2+\cos^2\frac{\theta}{2}-2y\cos\frac{\theta}{2}\sin\tilde\varphi\right] \nonumber\\
&=-2\pi \left(\cos^2\frac{\theta}{2}+\frac{1}{3}\sin^2\frac{\theta}{2}\right), \label{tildeI3final}
\end{align}
where we used the relation $l_2(\tilde\varphi)=l_2(-\tilde\varphi)$.

Using integrals $\tilde I_1$ (\ref{tildeI1final}), $\tilde I_2$ (\ref{tildeI2final}), and $\tilde I_3$ (\ref{tildeI3final}) in Eq.~(\ref{F12initial}), we obtain equation (\ref{tildeF12}):
\begin{equation*}
\!\!\!\tilde F_{1}^{(2)}(\theta)\!\!
=\!\!\frac{2\hbar^2r^2_*k^2_F}{m}\!\!\left[\! 3\pi\! +\!2\pi\sin^2\frac{\theta}{2}\left(\frac{4}{3} \!-\!\ln|\tan \frac{\theta}{2}|\right)\!\right].
\end{equation*}
.
\section{Calculation of the interaction function $\tilde F_{2}^{(2)}$} 
\label{F22}
The interaction function  $\tilde F_{2}^{(2)}$ is the second variational derivative of the many-body contribution to the interaction energy, $\tilde E^{(2)}_2$ (\ref{tildeE22}), with respect to the momentum distribution. It reads:
\begin{equation}    \label{F22initial}
\tilde F_{2}^{(2)}(\vec{k},\vec{k'})=\frac{2\hbar^2}{m}(k_Fr_*)^2(I'_1+I'_2), 
\end{equation}
where
\begin{align} 
\!\!&I'_1=2\!\int_{|\vec{k_1}|<k_F}\!\frac{d^2k_1}{k_F^2}\frac{|\vec{k}\!-\!\vec{k_1}|\cdot|\vec{k'}\!-\!\vec{k_1}|}{\vec{k^2}\!+\!\vec{k'^2}\!-\!\vec{k^2_1}\!-\!\vec{k^2_2}}\delta_{\vec{k}\!+\!\vec{k'}\!-\!\vec{k_1}\!-\!\vec{k_2}},    \label{I1prime} \\
\!\!&I'_2=4\!\int_{|\vec{k_1}|<k_F}\!\frac{d^2k_1}{k_F^2}\frac{|\vec{k}\!-\!\vec{k'}|\cdot|\vec{k_1}\!-\!\vec{k'}|}{\vec{k^2}\!+\!\vec{k^2_1}\!-\!\vec{k'^2}\!-\!\vec{k^2_2}}\delta_{\vec{k}\!+\!\vec{k_1}\!-\!\vec{k'}\!-\!\vec{k_2}} \label{I2prime}
\end{align}

\begin{figure}[ttp]
\includegraphics[width=0.5\columnwidth]{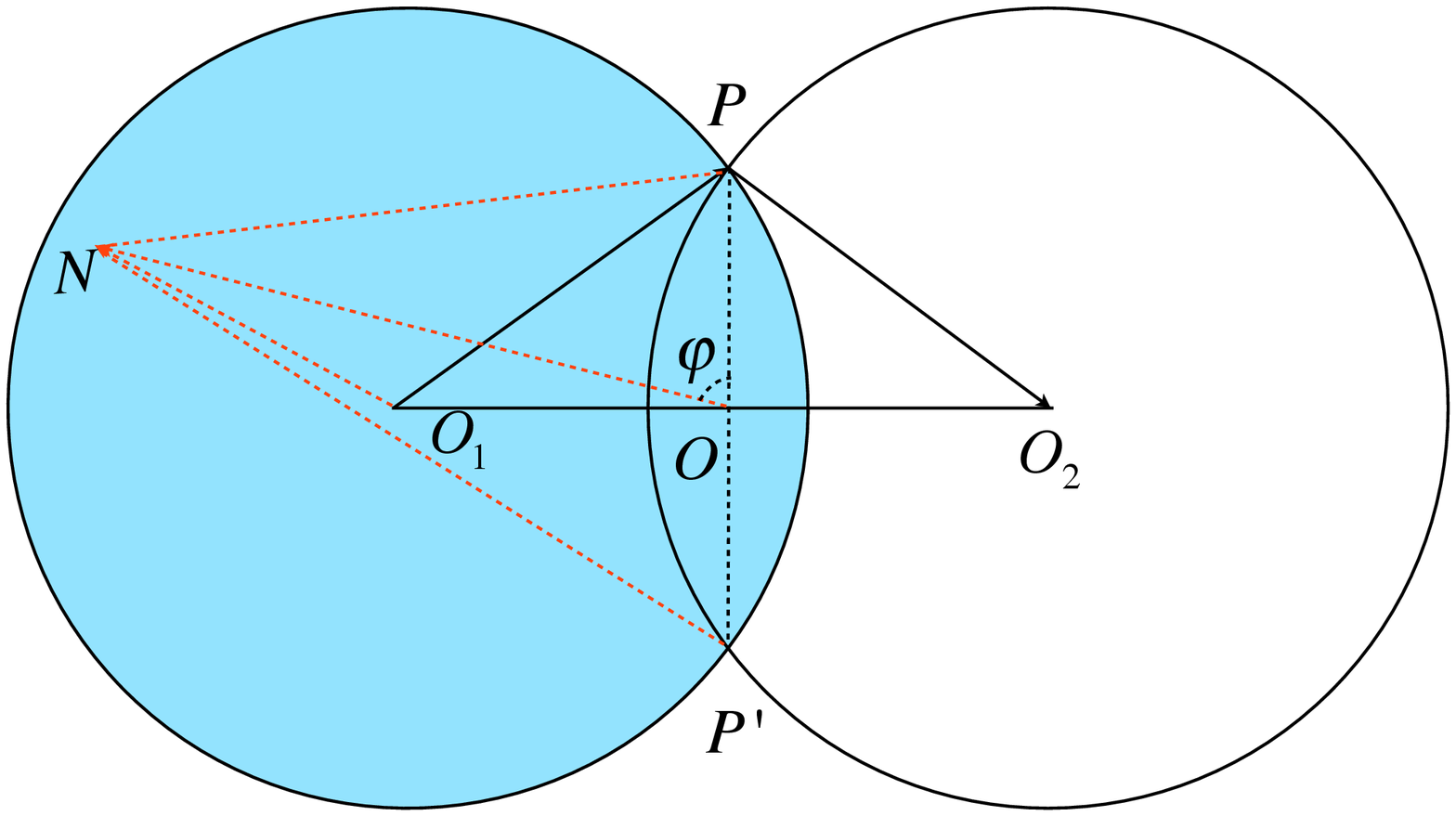}\hfill
\includegraphics[width=0.5\columnwidth]{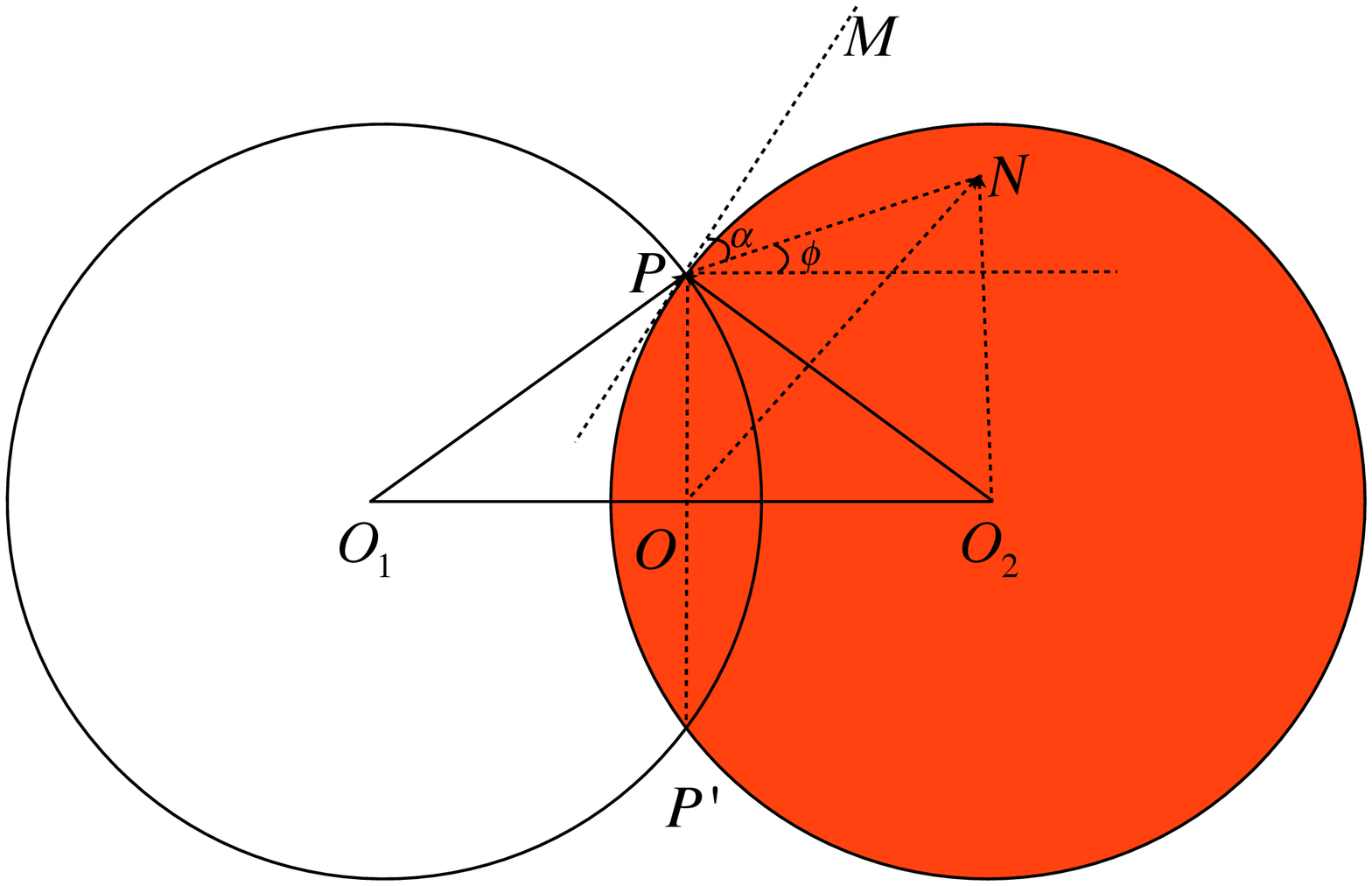}
\caption{(color online).
Left: The integration area for $I'_1$ (in blue): ${\bf R}_{O_1P}=\vec{k}/2k_F$, ${\bf R}_{PO_2}=\vec{k'}/2k_F$, ${\bf R}_{O_1N}=\vec{k_1}/2k_F$, and $\varphi$ is the angle between the vectors ${\bf R}_{OP}$ and ${\bf R}_{ON}$, which is the same as the angle between ${\bf m}$ and ${\bf x}$. Right: The integration area for $ I'_2$ (in red): ${\bf R}_{O_1P}=\vec{k}/2k_F$, ${\bf R}_{O_2P}=\vec{k'}/2k_F$, 
${\bf R}_{O_2N}=\vec{k_1}/2k_F$, $\alpha$ is the angle between ${\bf R}_{PM}$ and ${\bf R}_{PN}$, and $\phi$ is the angle between ${\bf R}_{PN}$ and ${\bf R}_{OO_2}$.
\label{fig:f2}}
\end{figure}

The integration area for $I'_1$ is shown in Fig.~\ref{fig:f2}, where the distance between the points $O_1$ and $P$ is ${\bf R}_{O_1P}={\bf k}/2k_F$, ${\bf R}_{PO_2}={\bf k}'/2k_F$, ${\bf R}_{O_1N}={\bf k}_1/2k_F$, and ${\bf R}_{ON}={\bf x}/2$. We thus have ${\bf R}_{NP}=({\bf k}-{\bf k}_1)/2k_F$ and ${\bf R}_{NP'}=({\bf k}'-{\bf k}_1)/2k_F$. In the region of integration we should have $|{\bf R}_{O_1N}|=k_1/2k_F\leq 1/2$. This leads to 
\begin{align}
I'_1&=4\int \frac{|{\bf R}_{NP}|\cdot|{\bf R}_{NP'}|}{m^2-x^2}d^2n=-\int_0^{2\pi} d\varphi\int_{0}^{l_3(\varphi)}xdx \nonumber \\
&\times\frac{\sqrt{[x^2+\sin^2(\theta/2)]^2-4x^2\sin^2(\theta/2)\cos^2\varphi}}{x^2-\sin^2(\theta/2)},
\end{align}
where $\varphi$ is the angle between ${\bf x}$ and ${\bf m}$ (see Fig.~\ref{fig:f2}), and the quantity $l_3(\varphi)$ obeys the equation
$$l_3^2(\varphi)-2\cos\frac{\theta}{2}\sin\varphi \cdot l_3(\varphi)+\cos^2\frac{\theta}{2}=1.$$  
Turning to the variable $z=r^2-\sin^2(\theta/2)$ the integral $I'_1$ is reduced to
\begin{align}
\label{I1primeinterm}
I'_1=-\frac{1}{2}\int_0^{2\pi} d\varphi \int_{-\sin^2(\theta/2)}^{l_3^2(\varphi)-\sin^2(\theta/2)} \frac{\sqrt{R}}{z} dz,
\end{align}
with
\begin{equation*}
R=z^2+4z\sin^2\frac{\theta}{2}\sin^2\varphi+4\sin^4\frac{\theta}{2}\sin^2\varphi.
\end{equation*}

It is easy to see that:
\begin{align*}
&I_r=\int_{-\sin^2(\theta/2)}^{l_3^2(\varphi)-\sin^2(\theta/2)}\frac{\sqrt{R}}{z}dz  \nonumber \\
&=\Big\{\sqrt{R}-\sqrt{a}\ln\left(2a+bz+2\sqrt{aR}\right)\nonumber \\
&+\frac{b}{2}\ln\left(2\sqrt{R}+2z+b\right)\Big\}\Big|^{l_3^2(\varphi)-\sin^2(\theta/2)}_{-\sin^2(\theta/2)} \nonumber \\
&+\sqrt{a} \cdot P\int_{-\sin^2(\theta/2)}^{l_3^2(\varphi)-\sin^2(\theta/2)}\frac{dz}{z}=I_{r\uparrow}-I_{r\downarrow},
\end{align*}
where $a=4\sin^4(\theta/2)\sin^2\varphi$, $b=4\sin^2(\theta/2)\sin^2\varphi$, and the symbol $P$ stands for the principal value of the integral. The quantities $I_{r\uparrow}$ and $I_{r\downarrow}$ denote the values of the integral at the upper and lower bounds, respectively (in the last line we have to take the principal value of the integral and, hence, if the upper bound of the integral is positive we have to replace the lower bound with $\sin^2(\theta/2)$). Then $I_{r\uparrow}$ and $I_{r\downarrow}$ are given by:
\begin{widetext}
\begin{align}
&I_{r\uparrow}=2|\sin\varphi|\cdot l(\varphi)-2\sin^2\frac{\theta}{2}|\sin\varphi|\cdot\left[\ln\left(8\sin^2\frac{\theta}{2}\sin^2\varphi\right)+\ln\left(\sin^2\frac{\theta}{2}+\cos\frac{\theta}{2}\sin\varphi \cdot l(\varphi)+l(\varphi)\right)\right] \nonumber\\
&+2\sin^2\frac{\theta}{2}|\sin\varphi|\cdot\ln|2\cos\frac{\theta}{2}\sin\varphi \cdot l(\varphi)|+2\sin^2\frac{\theta}{2}\sin^2\varphi\left[\ln4+\ln\left(|\sin\varphi|\cdot l(\varphi)+\cos\frac{\theta}{2}\sin\varphi\cdot l(\varphi)+\sin^2\frac{\theta}{2}\sin^2\varphi\right)\right], \nonumber \\
&I_{r\downarrow}=\sin^2\frac{\theta}{2}-2\sin^2\frac{\theta}{2}|\sin\varphi|\cdot\left[\ln\left(4\sin^4\frac{\theta}{2}\right)+\ln\left(\sin^2\varphi+|\sin\varphi|\right)\right]+2\sin^2\frac{\theta}{2}|\sin\varphi|\cdot\ln\left(\sin^2\frac{\theta}{2}\right)\nonumber\\
&+2\sin^2\frac{\theta}{2}\sin^2\varphi\cdot\ln\left(4\sin^2\frac{\theta}{2}\sin^2\varphi\right). \nonumber
\end{align}
\end{widetext}
The integral $I'_1$ can be expressed as: 
\begin{equation*}
I'_1=-\frac{1}{2}\int_{0}^{2\pi}[I_{r\uparrow}-I_{r\downarrow}]d\varphi,
\end{equation*}
and for performing the calculations we notice that $l_3(\varphi)\cdot l_3(\varphi+\pi)=\sin^2\frac{\theta}{2}$, $l_3(\varphi)-l_3(\varphi+\pi)=2\cos\frac{\theta}{2}\sin\varphi$, and 
$l_3(\varphi)+l_3(\varphi+\pi)=2\sqrt{\cos^2\frac{\theta}{2}\sin^2\varphi+\sin^2\frac{\theta}{2}}$. We then obtain:
\begin{widetext}
\begin{align}
\label{I11}
I'_{1}=&-\sin^2\frac{\theta}{2}\left(\pi\ln2+\pi/2-\pi\ln\sin\frac{\theta}{2}+4\ln|\cos\frac{\theta}{2}|-4\ln(1+\sin\frac{\theta}{2})+\mathcal{G}(\theta)-\frac{2\pi}{|\cos\frac{\theta}{2}|}-\frac{4\arcsin(\sin\frac{\theta}{2})}{|\cos\frac{\theta}{2}|} \right) \nonumber\\
&-\frac{k^2_F}{|\cos\frac{\theta}{2}|}\left(\pi-2\arcsin(\sin\frac{\theta}{2})+|\sin\theta|\right),
\end{align}
with
\begin{equation}
\mathcal{G}(\theta)=\int_{0}^{\pi} 2\sin^2\varphi \ln\left(\sin\varphi+\sqrt{\sin^2\frac{\theta}{2}+\cos^2\frac{\theta}{2}\sin^2\varphi}\right) d\varphi.
\end{equation}
\end{widetext}

The integration area for $I'_2$ is shown in Fig.~\ref{fig:f2}, and we get:
\begin{equation}
I'_2=-4\int \frac{|\vec{m}|\cdot|{\bf R}_{PN}|}{\vec{m}\cdot\vec{y}}d^2y=-8\int \frac{d^2\rho}{\cos \phi},
\end{equation}
where we denote ${\bf R}_{PN}=$ $\boldsymbol\rho$, and $\phi=\alpha-\theta/2$ is the angle between the vectors $\vec{m}$ and $\boldsymbol\rho$, with $\alpha$ being the angle between the vectors ${\bf R}_{PM}$  and ${\bf R}_{PN}$ (see Fig.~\ref{fig:f2}). We then have:
\begin{align}
\label{I12}
I'_2&=-8\int_{0}^{\pi} d\alpha\int_{0}^{\sin\alpha}\frac{\rho d\rho}{\cos (\alpha-\theta/2)}\nonumber\\
&=-4\left[\cos^2\frac{\theta}{2}\ln\frac{1+\sin(\theta/2)}{1-\sin(\theta/2)}+2\sin\frac{\theta}{2}\right].
\end{align}

Using $I'_1$~(\ref{I11}) and $I'_2$~(\ref{I12}) in Eq.~(\ref{F22initial}) we get equation (\ref{tildeF22}) for the interaction function $\tilde F^{(2)}_2(\theta)$. 


\begin{thebibliography}{10}
\bibitem{comment} In fact, we arrived at this conclusion confining ourselves to the contributions up to $\beta^2$. In stead of considering higher order contributions we checked numerically that the solution with $s>1$ is absent if we omit the $(k_Fr_*)^2$-terms in the interaction function of quasiparticles.
\end{thebibliography}
\end{document}